\documentclass[12pt]{article}
\usepackage[margin=1in]{geometry}
\usepackage{graphicx}
\graphicspath{ {./Figures/} }
\usepackage{float}
\usepackage[T1]{fontenc}
\usepackage[utf8]{inputenc}
\usepackage{mathtools}
\usepackage{enumitem}
\usepackage{amsmath}
\DeclareMathOperator*{\argmax}{arg\,max}
\usepackage{apacite}
\usepackage{natbib}
\usepackage[colorlinks, allcolors=blue]{hyperref}
\usepackage{enumitem}
\usepackage{bbm}
\usepackage{amssymb}
\usepackage{commath}
\usepackage{amsthm,amssymb}

\title{Mode Treatment Effect}
\author{Neng-Chieh Chang\footnote{Department of Economics, University of California Los Angeles, 315 Portola Plaza, Los Angeles, CA 90095, USA. email: nengchiehchang@g.ucla.edu}}
\date{}

\begin{document}
\maketitle

\begin{abstract}
Mean, median, and mode are three essential measures of the centrality of probability distributions. In program evaluation, the average treatment effect (mean) and the quantile treatment effect (median) have been intensively studied in the past decades. The mode treatment effect, however, has long been neglected in program evaluation. This paper fills the gap by discussing both the estimation and inference of the mode treatment effect. I propose both traditional kernel and machine learning methods to estimate the mode treatment effect. I also derive the asymptotic properties of the proposed estimators and find that both estimators follow the asymptotic normality but with the rate of convergence slower than the regular rate $\sqrt{N}$, which is different from the rates of the classical average and quantile treatment effect estimators.
\end{abstract}

\section{Introduction}
The effects of policies on the distribution of outcomes have long been of central interest in many areas of empirical economics. A policy maker might be interested in the difference of the distribution of outcome under treatment and the distribution of outcome in the absence of treatment. The empirical studies of distributional effects include but not are not limited to \citet*{freeman1980unionism}, \citet*{card1996effect}, \citet*{dinardo1995labor}, and \citet*{bitler2006mean}. Most researches use the difference of the averages or quantiles of the treated and untreated distribution, known as average treatment effect and quantile treatment effect, as a summary for the effect of treatment on distribution. The mode of a distribution, which is also an important summary statistics of data, has long been ignored in the literature. This paper fills up the gap by studying the mode treatment effect: the difference of the modes of the treated and untreated distribution. Compared to the average and the quantile treatment effect, the mode treatment effect has two advantages: (1) mode captures the most probable value of the distribution under treatment and in the absence of treatment. It provides a better summary of centrality than average and quantile when the distributions are highly skewed; (2) mode is robust to heavy-tailed distributions where outliers don't follow the same behavior as the majority of a sample. In economic studies, it is especially often to confront a skewed and heavy-tailed distribution when the outcome of interest is income or wage. 

This paper discusses the estimation and inference of the mode treatment effect under the Strong Ignorability assumption \citep*{rosenbaum1983central}, which states that conditional on a vector of control variables the treatment is randomly assigned. The first estimator I propose is the kernel estimator. I estimate the density function of the outcome distribution using the kernel method and define the maximum of the estimated density function as the estimator of the mode. While the kernel estimator is a straightforward estimator, it requires to estimate the conditional density function in the process, and the estimation of the conditional density function may be difficult in practice when there exist more than two or three control variables, due to the curse of dimension. The kernel estimator is appropriate if there are less than three control variables. In practice, however, researchers may want to include as many control variables as possible in order to make their identification robust. In this circumstance, the curse of dimension may lead to inaccurate estimation and misleading inference. 

To address this problem, I propose the ML estimator. The key feature of the proposed ML estimator is that it translates the estimation of the conditional density function into the estimation of conditional expectation, which we can apply a rich set of ML methods, such as Lasso, random forests, neural nets, and etc, to estimate. This feature provides researchers with the flexibility to apply ML methods to estimate the density function of the outcome distribution. By the virtue of ML methods, the proposed ML estimator can handle the situation when there exist many control variables, even the number of control variables is comparable to or more than the sample size. However, it is well-known that the regularization bias embedded in ML methods may lead to the bias of the final estimator and misleading inference \citep*{chernozhukov2018double}. To solve this problem, I further derive the Neyman-orthogonal scores \citep*{chernozhukov2018double} for each estimation which requires the first-step estimation of the conditional expectation. These Neyman-orthogonal scores, to my best knowledge, are new results. The proposed ML estimator is built on the newly derived Neyman-orthogonal score, and hence, it is robust to the regularization bias of the first-step ML estimation. 

I derive the asymptotic properties for both the proposed kernel and ML estimators. I show that both estimators are consistent and asymptotically normal with the rate of convergence $\sqrt{Nh^{3}}$, where $N$ is the sample size and $h$ is the bandwidth of the chosen kernel, which is slower than the traditional rate of convergence $\sqrt{N}$ in the estimation of mean and quantile. In fact, this rate of convergence complies the intuition. While the estimators of mean and quantile are the weighted average of all the available observations, only a small portion of observations near the mode provides the information to the estimator of the mode. This explains the slower rate of convergence for the proposed estimators. 

This paper contributes to the program evaluation literature which includes the studies of average treatment effect: \citet*{rosenbaum1983central}, \citet*{heckman1985alternative}, \citet*{heckman1997matching},  \citet*{hahn1998role}, and \citet*{hirano2003efficient}; the studies of quantile treatment effect: \citet*{abadie2002instrumental}, \citet*{chernozhukov2005iv}, and \citet*{firpo2007efficient}; the studies of mode estimation and mode regression: \citet*{parzen1962estimation}, \citet*{eddy1980optimum}, \citet*{lee1989mode},  \citet*{yao2014new}, and \citet*{chen2016nonparametric}; as well as the causal inference of ML methods: \citet{belloni2012sparse}, \citet{Belloni14restud}, \citet{chernozhukov2015valid}, \citet{belloni2017program}, \citet{chernozhukov2018double}, and \citet{athey2019generalized}. This paper is also closely related to the robustness of average treatment effect estimation discussed in \citep*{robins1995semiparametric} and the general discussion in \citep*{chernozhukov2016locally}. The asymptotic properties of the robust estimators discussed in these papers remain unaffected if only one of the first-step estimation with classical nonparametric method is inconsistent.

\textbf{Plan of the paper.} Section 2 sets up the notation and framework for the discussion of the mode treatment effect. Section 3 discusses the kernel method and derives the asymptotic properties. Section 4 presents the ML  estimator for density estimation and the corresponding Neyman-orthogonal score. I combine the Neyman-orthogonal score with the cross-fitting algorithm to propose the ML estimator of the mode treatment effect, and derive its asymptotic properties. Section 5 concludes this paper.

\section{Notation and Framework}
Let Y be a continuous outcome variable of interest, D the binary treatment indicator, and X $d\times 1$ vector of control variables. Denote by $Y_{1}$ an individual's potential outcome when $D=1$ and $Y_{0}$ if $D=0$. Let $f_{Y_{1}}\left(y\right)$ and $f_{Y_{0}}\left(y\right)$ be the marginal probability density function (p.d.f.) of $Y_{1}$ and $Y_{0}$, respectively. The modes of $Y_{1}$ and $Y_{0}$ are the values that appear with the highest probability. That is, 
\[
\theta_{1}^{*}\equiv\argmax_{y\in \mathcal{Y}_{1}}f_{Y_{1}}\left(y\right)\text{ and }\theta_{0}^{*}\equiv\argmax_{y\in \mathcal{Y}_{0}}f_{Y_{0}}\left(y\right),
\]  
where $\mathcal{Y}_{1}$ and $\mathcal{Y}_{0}$ are the supports of $Y_{1}$ and $Y_{0}$. Here I assume that $\theta_{1}^{*}$ and $\theta_{0}^{*}$ are unique, meaning that both $Y_{1}$ and $Y_{0}$ are unimodal. I also assume that the modes $\theta_{1}^{*}$ and $\theta_{0}^{*}$ are in the interior of the common supports of $Y_{1}$ and $Y_{0}$. These conditions are formally stated in the following assumption:

\begin{flushleft}
\textbf{Assumption 1 (Uni-mode)}
\begin{itemize}
\item \textit{For all} $\epsilon>0$, 
\[
\sup_{y:\mid y-\theta_{1}^{*}\mid>\varepsilon}f_{Y_{1}}\left(y\right)<f_{Y_{1}}\left(\theta_{1}^{*}\right)\text{ for } y\in\mathcal{Y}_{1},
\]
\textit{and}
\[
\sup_{y:\mid y-\theta_{0}^{*}\mid>\varepsilon}f_{Y_{0}}\left(y\right)<f_{Y_{0}}\left(\theta_{0}^{*}\right)\text{ for } y\in\mathcal{Y}_{0}.
\]
\item $\theta_{1}^{*}, \theta_{0}^{*}\in Int\left( \mathcal{Y}_{1}\cap \mathcal{Y}_{0}\right)$.
\end{itemize}
\end{flushleft}
Assumption 1 has been widely adopted in many studies \citep*{parzen1962estimation, eddy1980optimum, lee1989mode, yao2014new}. Under Assumption 1, the mode treatment effect is uniquely defined as $\Delta^{*}\equiv\theta_{1}^{*}-\theta_{0}^{*}$. The following states the strong ignorability assumption \citep*{rosenbaum1983central}:

\begin{flushleft}
\textbf{Assumption 2 (strong ignorability)}
\begin{itemize}
\item  $\left(Y_{0},Y_{1}\right)\perp D\mid X$
\item  $0<P\left(D=1\mid X\right)<1$
\end{itemize}

\end{flushleft}
The first part of Assumption 2 assumes that potential outcomes are independent of treatment after conditioning on the observable covariates $X$. The second part states that for all values of $X$, both treatment status occur with a positive probability. Under the strong ignorability condition, both $f_{Y_1}$ and $f_{Y_0}$ can be identified from the observable variables $\left(Y,D,X\right)$ since

\begin{align*}
f_{Y\mid D=1,X}\left(y\mid x\right) = f_{Y_{1}\mid D=1,X}\left(y\mid x\right)=f_{Y_{1}\mid X}\left(y\mid x\right),
\end{align*}
and thus 
\[
f_{Y_{1}}(y)=E\left[ f_{Y_{1}\mid X}\left(y\mid X\right)\right]=E\left[ f_{Y\mid D=1,X}\left(y\mid X\right)\right].\tag{2.1}
\]
Similarly, we have 
\[
f_{Y_{0}}\left(y\right)=E\left[ f_{Y\mid D=0,X}\left(y\mid X\right)\right].\tag{2.2}
\] 
Equation (2.1) and (2.2) shows the identification result of the density function $f_{Y_{1}}$ and $f_{Y_{0}}$. Then it is straightforward to identify their modes $\theta_{1}^{*}$ and $\theta_{0}^{*}$:
\[
\theta_{1}^{*}=\argmax_{y\in\mathcal{Y}_{1}}E\left[ f_{Y\mid D=1,X}\left(y\mid X\right)\right]\text{ and }\theta_{0}^{*}=\argmax_{y\in\mathcal{Y}_{0}}E\left[ f_{Y\mid D=0,X}\left(y\mid X\right)\right].\tag{2.3}
\]
If both $f_{Y\mid D=1,X}\left(y\mid X\right)$ and $f_{Y\mid D=0,X}\left(y\mid X\right)$ are differentiable with respect to $y$, we can further identify the modes using the first-order conditions under Assumption 1:
\[
E\left[ f_{Y\mid D=1,X}^{\left(1\right)}\left(\theta_{1}^{*}\mid X\right)\right]=0\text{ and }E\left[ f_{Y\mid D=0,X}^{\left(1\right)}\left(\theta_{0}^{*}\mid X\right)\right]=0,\tag{2.4}
\]
where $m^{\left(s\right)}\left(y, x\right)\equiv \partial^{s} m\left(y, x\right)/ \partial y^{s}$ denotes the partial derivatives with respect to $y$.

Equation (2.1)-(2.4) provide us a direct way to estimate the modes $\theta_{1}^{*}$ and $\theta_{0}^{*}$. Intuitively, we estimate the density functions $f_{Y_{1}}(y)$ and $f_{Y_{0}}(y)$ in the first step and use the maximizers of the estimated density functions as the estimators of the modes. Section 3 and 4 presents the kernel and ML estimation method, respectively. 

\section{The Kernel Estimation}
In this section, I propose kernel estimators for $\theta_{1}^{*}$, $\theta_{0}^{*}$, and the mode treatment effect $\Delta^{*}=\theta_{1}^{*}-\theta_{0}^{*}$. Let $K(\cdot)$ be a kernel function with bandwidth $h$. Define the estimators of the density functions $f_{Y_{1}}\left(y\right)$ and $f_{Y_{0}}\left(y\right)$ as,
\[
\hat{f}_{Y_{1}}\left(y\right)=\frac{1}{n}\sum_{i=1}^{n}\hat{f}_{Y\mid D=1,X}\left(y\mid X_{i}\right),
\]
\[
\hat{f}_{Y_{0}}\left(y\right)=\frac{1}{n}\sum_{i=1}^{n}\hat{f}_{Y\mid D=0,X}\left(y\mid X_{i}\right)
\]
with the kernel estimators
\[
\hat{f}_{Y\mid D=1,X}\left(y\mid x\right) = \frac{\sum_{j=1}^{n}D_{j}K_{h}\left(y-Y_{j}\right)K_{h}\left(x-X_{j}\right)}{\sum_{j=1}^{n}D_{j}K_{h}\left(x-X_{j}\right)},
\]
\[
\hat{f}_{Y\mid D=0,X}\left(y\mid x\right) = \frac{\sum_{j=1}^{n}\left(1-D_{j}\right)K_{h}\left(y-Y_{j}\right)K_{h}\left(x-X_{j}\right)}{\sum_{j=1}^{n}\left(1-D_{j}\right)K_{h}\left(x-X_{j}\right)}
\]
where $K_{h}\left(y-Y_{j}\right)= h^{-1}K\left(\frac{y-Y_{j}}{h}\right)$ and
\[
K_{h}\left(x-X_{j}\right)= h^{-d}K\left(\frac{x_{1}-X_{j1}}{h}\right)\times ... \times K\left(\frac{x_{d}-X_{jd}}{h}\right).
\]

Then it is straightforward to define the estimators of the modes $\theta_{1}^{*}$ and $\theta_{0}^{*}$:

\[
\hat{\theta}_{1}\equiv \argmax_{y} \hat{f}_{Y_{1}}\left(y\right),
\]
\[
\hat{\theta}_{0}\equiv\argmax_{y} \hat{f}_{Y_{0}}\left(y\right).
\]
The estimator of the mode treatment effect $\Delta^{*}$ is $\hat{\Delta}\equiv \hat{\theta}_{1}-\hat{\theta}_{0}$. Through out the paper, I impose the following conditions on the kernel $K\left(\cdot\right)$:

\begin{flushleft}
\textbf{Assumption 3:}
\begin{itemize}
\item $\abs{K\left(u\right)}\leq \bar{K}<\infty$.
\item $\int K\left(u\right)du=1, \int uK\left(u\right)du=0, \int u^{2}K\left(u\right)du<\infty$.
\item $K\left(u\right)$ is differentiable. 
\end{itemize}
\end{flushleft} 
The first part of Assumption 3 requires that $K\left(u\right)$ is bounded. Although the second part implies that $K\left(u\right)$ is a first-order kernel, the arguments in this paper can be easily extended to higher-order kernels. We assume the first-order kernel here just for simplicity. The third part imposes enough smoothness on $K\left(u\right)$.

\begin{flushleft}
\textbf{Theorem 1 (Consistency)} \textit{Suppose Assumption 1-3 hold. Assume that the density functions $f_{Y\mid D=1,X}\left(y\mid x\right)$ and $f_{Y\mid D=0,X}\left(y\mid x\right)$ are (i) continuous in $y$, (ii) bounded by some function $d\left(x\right)$ with $E\left[d\left(X\right)\right]< \infty$ for all $y\in \mathcal{Y}$, and (iii) $y\in \mathcal{Y}$ and $x\in \mathcal{X}$ with compact $\mathcal{Y}$ and $\mathcal{X}$. We also assume that the density functions $f_{X\mid D=1}\left(x\right)$ and $f_{X\mid D=0}\left(x\right)$ are bounded away from zero. If $n\to \infty$, $h\to 0$, and $\ln{n} \left(nh^{d+1}\right)^{-1}\to  0$, then we have $\hat{\theta}_{1}\overset{p}{\to} \theta_{1}^{*}$ and $\hat{\theta}_{0}\overset{p}{\to} \theta_{0}^{*}$.}
\end{flushleft}

\begin{flushleft}
\textbf{Theorem 2 (Asymptotic Normality)}\textit{ Suppose that the assumptions of Theorem 1 hold. Assume that $f_{Y\mid X,D=1}^{\left(2\right)}\left(y\mid x\right)$ and $f_{Y\mid X,D=0}^{\left(2\right)}\left(y\mid x\right)$ are continuous at $y=\theta_{1}^{*}$ and $y=\theta_{0}^{*}$ for all $x$, respectively. If $n\to \infty$, $h\to 0$, $\sqrt{nh^{3}}\left(\ln{n}\right)\left(nh^{d+3}\right)^{-1}\to 0$, $\left(\ln{n}\right)\left(nh^{d+5}\right)^{-1}\to 0$, and $\sqrt{nh^{3}}h^{2}\to 0$, then }
\[
\sqrt{nh^{3}}\left(\hat{\theta}_{1}-\theta_{1}^{*}\right)\overset{d}\to N\left(0,M_{1}^{-1}V_{1}M_{1}^{-1}\right)
\]
\[
\sqrt{nh^{3}}\left(\hat{\theta}_{0}-\theta_{0}^{*}\right)\overset{d}\to N\left(0,M_{0}^{-1}V_{0}M_{0}^{-1}\right)
\]
\textit{where} $$M_{1}\equiv E\left[f_{Y\mid X,D=1}^{\left(2\right)}\left(\theta_{1}^{*}\mid X\right)\right],$$ 

$$M_{0}\equiv E\left[f_{Y\mid X,D=0}^{\left(2\right)}\left(\theta_{0}^{*}\mid X\right)\right],$$

$$V_{1}=\kappa_{0}^{\left(1\right)}E\left[\frac{f_{Y\mid X, D=1}\left(\theta_{1}^{*}\mid X\right)}{P\left(D=1\mid X\right)}\right],$$ 

$$V_{0}=\kappa_{0}^{\left(1\right)}E\left[\frac{f_{Y\mid X, D=0}\left(\theta_{0}^{*}\mid X\right)}{P\left(D=0\mid X\right)}\right],$$ 

and $\kappa_{0}^{\left(1\right)}=\int K^{\left(1\right)}\left(u\right)^{2}du$. Further, we have
\[
\sqrt{nh^{3}}\left(\hat{\Delta}-\Delta^{*}\right)\overset{d}{\to}N\left(0,M_{1}V_{1}M_{1}+M_{0}V_{0}M_{0}\right).
\]
\end{flushleft} 

Theorem 1 and 2 show that the asymptotic properties of the estimator of the mode treatment effect. We can see that the proposed estimators follows the asymptotic normality but with the rate of convergence slower than the regular rate $\sqrt{N}$. The intuition is that, unlike the estimation of the average and the quantile treatment effect, the estimation of modes only uses a small portion of total observations which are around the modes. The usage rate of observations determines that the rate of convergence is slower than the regular rate $\sqrt{N}$. 

To estimate the asymptotic variances, we define $\pi_{0}\left(X\right)\equiv P\left(D=1\mid X\right)$ to be the propensity score. The consistent variance estimators are
\[
\hat{M}_{1}=\frac{1}{n}\sum_{i=1}^{n}\hat{f}_{Y\mid X, D=1}^{\left(2\right)}\left(\hat{\theta}_{1}\mid X_{i}\right),
\]
\[
\hat{M}_{0}=\frac{1}{n}\sum_{i=1}^{n}\hat{f}_{Y\mid X, D=0}^{\left(2\right)}\left(\hat{\theta}_{0}\mid X_{i}\right),
\]
\[
\hat{V}_{1}=\kappa_{0}^{\left(1\right)}\frac{1}{n}\sum_{i=1}^{n}\frac{\hat{f}_{Y\mid X,D=1}\left(\hat{\theta}_{1}\mid X_{i}\right)}{\hat{\pi}\left(X_{i}\right)},
\] 
\[
\hat{V}_{0}=\kappa_{0}^{\left(1\right)}\frac{1}{n}\sum_{i=1}^{n}\frac{\hat{f}_{Y\mid X,D=0}\left(\hat{\theta}_{0}\mid X_{i}\right)}{\hat{\pi}\left(X_{i}\right)}.
\]
\begin{flushleft}
\textbf{Theorem 3 (Variance Estimation)} \textit{ Suppose that the assumptions in Theorem 2 hold. Let $\hat{\pi}\left(x\right)$ be an uniformly consistent estimator for $\pi_{0}\left(x\right)$. If $n\to \infty$, $h\to 0$, and $\ln{n}\left(nh^{d+5}\right)\to 0$, then $\hat{M}_{1}\overset{p}{\to} M_{1}$, $\hat{M}_{0}\overset{p}{\to} M_{0}$, $\hat{V}_{1}\overset{p}{\to} V_{1}$, $\hat{V}_{0}\overset{p}{\to} V_{0}$. Thus we have $\hat{M}_{1}^{-1}\hat{V}_{1}\hat{M}_{1}^{-1}\overset{p}{\to} M_{1}^{-1}V_{1}M_{1}^{-1}$ and $\hat{M}_{0}^{-1}\hat{V}_{0}\hat{M}_{0}^{-1}\overset{p}{\to} M_{0}^{-1}V_{0}M_{0}^{-1}.$}
\end{flushleft}

\section{The Machine Learning Estimation}
In this section, I propose the ML estimator of the mode treatment effect. The ML estimator can accommodate a large number of control variables, potentially more than the sample size. This flexibility will enable researcher to include as many control variables they consider important to make their identification assumptions more plausible. The key to implement ML methods is to replace the estimation of the conditional density function with the estimation of the conditional expectation. To begin with, the estimation of the conditional density function in the traditional kernel estimation is

\[
\hat{f}_{Y\mid D=1,X}\left(y\mid x\right) = \frac{\sum_{j=1}^{n}D_{j}K_{h}\left(y-Y_{j}\right)K_{h}\left(x-X_{j}\right)}{\sum_{j=1}^{n}D_{j}K_{h}\left(x-X_{j}\right)}.
\]
Notice that we can divide both the numerator and the denominator by $\sum_{j=1}^{n}K_{h}\left(x-X_{j}\right)$ to  obtain
\begin{align*}
\hat{f}_{Y\mid D=1,X}\left(y\mid x\right) & = \frac{\sum_{j=1}^{n}D_{j}K_{h}\left(y-Y_{j}\right)K_{h}\left(x-X_{j}\right)/\sum_{j=1}^{n}K_{h}\left(x-X_{j}\right)}{\sum_{j=1}^{n}D_{j}K_{h}\left(x-X_{j}\right)/\sum_{j=1}^{n}K_{h}\left(x-X_{j}\right)}.
\end{align*}
The numerator is an kernel estimator of $E\left[DK_{h}\left(y-Y\right)\mid X\right]$ and the denominator is an kernel estimator of the propensity score $E\left[D\mid X\right]=\pi\left(X\right)$.  Hence, $\hat{f}_{Y\mid D=1,X}\left(y\mid x\right)$ is an estimator of $E\left[DK_{h}\left(y-Y\right)\mid X\right]/\pi\left(X\right)$. Then the marginal density estimator 
\[
\hat{f}_{Y_{1}}\left(y\right)=\frac{1}{n}\sum_{i=1}^{n}\hat{f}_{Y\mid D=1,X}\left(y\mid X_{i}\right)
\]
defined in the previous section can be interpreted as an estimator of 
\[
E\left[\frac{E\left[DK_{h}\left(y-Y\right)\mid X\right]}{\pi\left(X\right)}\right]=E\left[\frac{DK_{h}\left(y-Y\right)}{\pi\left(X\right)}\right].
\]
Therefore, we can use the machine learning estimator of $E\left[\frac{DK_{h}\left(y-Y\right)}{\pi\left(X\right)}\right]$ as an estimator for $f_{Y_{1}}\left(y\right)$. We have successfully translate the estimation of the conditional density function into the estimation of the conditional expectation, which is the propensity score $\pi(X)$.

Here we pursue a little bit further to construct the Neyman-orthogonal   score \citep*{chernozhukov2018double} for the robustness of the first-step estimation: 
\[
m_{1}\left(Z, y, \eta_{10}\right)=\frac{DK_{h}\left(y-Y\right)}{\pi_{0}\left(X\right)}-\frac{D-\pi_{0}\left(X\right)}{\pi_{0}\left(X\right)}E\left[K_{h}\left(y-Y\right)\mid X, D=1\right],\tag{4.1}
\]
where $Z=\left(Y, D, X\right)$ and $\eta_{0}=\left(\pi_{0}, g_{10} \right)$ with $g_{10}\left(X\right)\equiv E\left[K_{h}\left(y-Y\right)\mid X, D=1\right]$. Similary, the Neyman-orthogonal score for $f_{Y_{0}}\left(y\right)$ is 

\[
m_{2}\left(Z, y, \eta_{20}\right)=\frac{\left(1-D\right)K_{h}\left(y-Y\right)}{1-\pi_{0}\left(X\right)}-\frac{\pi_{0}-D\left(X\right)}{1-\pi_{0}\left(X\right)}E\left[K_{h}\left(y-Y\right)\mid X, D=0\right], \tag{4.2}
\]
where $\eta_{20}=\left(\pi_{0}, g_{20}\right)$ with $g_{20}\left(X\right)\equiv E\left[K_{h}\left(y-Y\right)\mid X, D=0\right]$. Equation (4.1) and (4.2), to my best knowledge, should be the new results for density estimation. The Neyman orthogonality will make the estimation of the density functions more robust to the first-step estimation. Now I combine (4.1) and (4.2) with the cross-fitting algorithm \citep*{chernozhukov2018double} to propose the new estimator:
\begin{flushleft}
\textbf{Definition.} 
\end{flushleft}
\begin{enumerate}
\item Take a $K$-fold random partition $\left(I_{k}\right)_{k=1}^{K}$ of $\left[N\right]=\left\{1,...,N\right\}$ such that the size of each $I_{k}$ is $n=N/K$.  For each $k\in\left[K\right]=\left\{1,...,K\right\}$, define the auxiliary sample $I_{k}^{c}\equiv \left\{1,...,N\right\}$. 
\item For each $k\in\left[K\right]$, use the auxiliary sample $I_{k}^{c}$ to construct machine learning estimators 
\[
\hat{\pi}_{k}\left(x\right)\text{, }\hat{g}_{1k}\left(x\right)\text{, and }\hat{g}_{2k}\left(x\right)
\]
of $\pi_{0}\left(x\right)$, $g_{10}\left(x\right)$, and $g_{20}\left(x\right)$.
\item Construct the estimator of $f_{Y_{1}}\left(y\right)$ and $f_{Y_{0}}\left(y\right)$:
\[
\hat{f}_{Y_{1}}\left(y\right)=\frac{1}{K}\sum_{k=1}^{K}\mathbb{E}_{n,k}\left[m_{1}\left(Z, y, \hat{\eta}_{1k}\right)\right]\text{ and }\hat{f}_{Y_{0}}\left(y\right)=\frac{1}{K}\sum_{k=1}^{K}\mathbb{E}_{n,k}\left[m_{2}\left(Z, y, \hat{\eta}_{2k}\right)\right]
\]
where $\mathbb{E}_{n,k}\left[m\left(Z\right)\right]=n^{-1}\sum_{i\in I_{k}}m\left(Z_{i}\right)$. 
\item Construct the estimator for $\theta_{1}^{*}$ and $\theta_{0}^{*}$

\[
\hat{\theta}_{1}=\argmax_{y}\hat{f}_{Y_{1}}\left(y\right)\text{ and }\hat{\theta}_{0}=\argmax_{y}\hat{f}_{Y_{0}}\left(y\right).
\]
\item Construct the estimator for the mode treatment effect $\hat{\Delta}=\hat{\theta}_{1}-\hat{\theta}_{0}$.
\end{enumerate}

\begin{flushleft}
\textbf{Theorem 4.1.} \textit{Suppose that with probability $1-o\left(1\right)$,
$\parallel\hat{\eta}_{1k}-\eta_{10}\parallel_{P,2}\leq\varepsilon_{N}$,
$\parallel\hat{\pi}_{k}-1/2\parallel_{P,\infty}\leq1/2-\kappa$,
and $\parallel\hat{\pi}_{k}-\pi_{0}\parallel_{P,2}^{2}+\parallel\hat{\pi}_{k}-\pi_{0}\parallel_{P,2}\times\parallel\hat{g}_{1k}-g_{10}\parallel_{P,2}\leq\left(\varepsilon_{N}\right)^{2}$. If $\epsilon_{N}=o((Nh^{3})^{-1/4})$ and $Nh^{7}\to 0$, then we have}
\end{flushleft}

\[
\sqrt{nh^{3}}\left(\hat{\theta}_{1}-\theta_{1}^{*}\right)\overset{d}\to N\left(0,M_{1}^{-1}V_{1}M_{1}^{-1}\right),
\]
\[
\sqrt{nh^{3}}\left(\hat{\theta}_{0}-\theta_{0}^{*}\right)\overset{d}\to N\left(0,M_{0}^{-1}V_{0}M_{0}^{-1}\right).
\]

As for the variance estimation, recall that the kernel estimator of $M_{1}$ in the previous section is $\hat{M}_{1}=N^{-1}\sum_{i=1}^{N}\hat{f}_{Y\mid D=1,X}^{(2)}(\hat{\theta}_{1}\mid x)$, where

\[
\hat{f}_{Y\mid D=1,X}^{(2)}\left(y\mid x\right) = \frac{\sum_{j=1}^{n}D_{j}K_{h}\left(y-Y_{j}\right)^{(2)}K_{h}\left(x-X_{j}\right)}{\sum_{j=1}^{n}D_{j}K_{h}\left(x-X_{j}\right)}.
\]
Notice that we can divide both the numerator and the denominator by $\sum_{j=1}^{n}K_{h}\left(x-X_{j}\right)$ to  obtain
\begin{align*}
\hat{f}_{Y\mid D=1,X}^{(2)}\left(y\mid x\right) & = \frac{\sum_{j=1}^{n}D_{j}K_{h}\left(y-Y_{j}\right)^{(2)}K_{h}\left(x-X_{j}\right)/\sum_{j=1}^{n}K_{h}\left(x-X_{j}\right)}{\sum_{j=1}^{n}D_{j}K_{h}\left(x-X_{j}\right)/\sum_{j=1}^{n}K_{h}\left(x-X_{j}\right)}.
\end{align*}
Observe that the numerator is an kernel estimator of $E\left[DK_{h}\left(y-Y\right)\mid X\right]$ and the denominator is an kernel estimator of the propensity score $E\left[D\mid X\right]=\pi\left(X\right)$.  Hence, $\hat{f}_{Y\mid D=1,X}\left(y\mid x\right)$ is an estimator of $E\left[DK_{h}^{(2)}\left(y-Y\right)\mid X\right]/\pi\left(X\right)$. Hence, we can use the machine learning estimator of $E\left[\frac{DK_{h}^{(2)}\left(y-Y\right)}{\pi\left(X\right)}\right]$ as an estimator for $M_{1}$. We can also construct a DML estimator using the Neyman-orthogonal functional form

\[
\frac{DK_{h}^{(2)}\left(y-Y\right)}{\pi\left(X\right)}-\frac{D-\pi_{0}\left(X\right)}{\pi_{0}\left(X\right)}E\left[K_{h}^{(2)}\left(y-Y\right)\mid X, D=1\right]
\]
In step 1, we use machine learning methods to estimate $\pi_{0}(X)$ and $E[K_{h}^{(2)}\left(\hat{\theta}_{1}-Y\right)\mid X, D=1]$ using auxiliary sample $I_{k}^{c}$. In Step 2, we construct the DML estimator of $M_{1}$:

\[
\hat{M}_{1}=\frac{1}{K}\sum_{k=1}^{K}\sum_{i\in I_{k}}\frac{D_{i}K_{h}^{(2)}\left(\hat{\theta}_{1}-Y_{i}\right)}{\hat{\pi}\left(X_{i}\right)}-\frac{D_{i}-\hat{\pi}_{0}\left(X_{i}\right)}{\hat{\pi}_{0}\left(X_{i}\right)}\hat{E}\left[K_{h}^{(2)}\left(\hat{\theta}_{1}-Y\right)\mid X_{i}, D=1\right].
\]
By the general DML theory \citep*{chernozhukov2018double}, $\hat{M_{1}}$ is a consistent estimator of $M_{1}$. Similarly, we can construct the DML estimators for $V_{1}$, $M_{0}$, and $V_{0}$ using the following table: 
\begin{center}
\begin{tabular}{ |c|c|c| } 
 \hline
  & Original Form & Equivalent Form \\ 
 \hline
 $M_{1}$ & $E\left[f_{Y\mid X,D=1}^{\left(2\right)}\left(\theta_{1}^{*}\mid X\right)\right]$ & $E\left[\frac{DK_{h}^{(2)}\left(\theta_{1}^{*}-Y\right)}{\pi\left(X\right)}-\frac{D-\pi_{0}\left(X\right)}{\pi_{0}\left(X\right)}E\left[K_{h}^{(2)}\left(y-Y\right)\mid X, D=1\right]\right]$ \\
 \hline 
 $V_{1}$ & $\kappa_{0}^{\left(1\right)}E\left[\frac{f_{Y\mid X, D=1}\left(\theta_{1}^{*}\mid X\right)}{P\left(D=1\mid X\right)}\right]$ & $E\left[\frac{DK_{h}\left(\theta_{1}^{*}-Y\right)}{\pi\left(X\right)^{2}}-2\frac{D-\pi_{0}\left(X\right)}{\pi_{0}^{2}\left(X\right)}E\left[K_{h}\left(y-Y\right)\mid X, D=1\right]\right]$ \\ 
 \hline 
 $M_{0}$ & $E\left[f_{Y\mid X,D=0}^{\left(2\right)}\left(\theta_{1}^{*}\mid X\right)\right]$ & $E\left[\frac{(1-D)K_{h}^{(2)}\left(\theta_{1}^{*}-Y\right)}{1-\pi\left(X\right)}-\frac{\pi_{0}(X)-D}{1-\pi_{0}\left(X\right)}E\left[K_{h}^{(2)}\left(y-Y\right)\mid X, D=0\right]\right]$ \\
 \hline 
 $V_{0}$ & $\kappa_{0}^{\left(1\right)}E\left[\frac{f_{Y\mid X, D=0}\left(\theta_{1}^{*}\mid X\right)}{P\left(D=0\mid X\right)}\right]$ & $E\left[\frac{(1-D)K_{h}\left(\theta_{1}^{*}-Y\right)}{(1-\pi\left(X\right))^{2}}-2\frac{\pi_{0}\left(X\right)-D}{(1-\pi_{0}\left(X\right))^{2}}E\left[K_{h}\left(y-Y\right)\mid X, D=0\right]\right]$ \\
 \hline
\end{tabular}
\end{center}

\section{Conclusion}
This paper studies the estimation and inference of the mode treatment effect, which has been ignored in the treatment effect literature compared to the estimation of the average and the quantile treatment effect estimation. I propose both kernel and ML estimators to accommodate a variety of data sets faced by researchers. I also derive the asymptotic properties of the proposed estimators. I show that both estimators are consistent and asymptotically normal with the rate of convergence $\sqrt{Nh^{3}}$.

\bibliographystyle{apacite}
\bibliography{References}

\begin{thebibliography}{}

\bibitem [\protect \citeauthoryear {%
Abadie%
, Angrist%
\BCBL {}\ \BBA {} Imbens%
}{%
Abadie%
\ \protect \BOthers {.}}{%
{\protect \APACyear {2002}}%
}]{%
abadie2002instrumental}
\APACinsertmetastar {%
abadie2002instrumental}%
\begin{APACrefauthors}%
Abadie, A.%
, Angrist, J.%
\BCBL {}\ \BBA {} Imbens, G.%
\end{APACrefauthors}%
\unskip\
\newblock
\APACrefYearMonthDay{2002}{}{}.
\newblock
{\BBOQ}\APACrefatitle {Instrumental variables estimates of the effect of
  subsidized training on the quantiles of trainee earnings} {Instrumental
  variables estimates of the effect of subsidized training on the quantiles of
  trainee earnings}.{\BBCQ}
\newblock
\APACjournalVolNumPages{Econometrica}{70}{1}{91--117}.
\PrintBackRefs{\CurrentBib}

\bibitem [\protect \citeauthoryear {%
Athey%
, Tibshirani%
, Wager%
\BCBL {}\ \protect \BOthers {.}}{%
Athey%
\ \protect \BOthers {.}}{%
{\protect \APACyear {2019}}%
}]{%
athey2019generalized}
\APACinsertmetastar {%
athey2019generalized}%
\begin{APACrefauthors}%
Athey, S.%
, Tibshirani, J.%
, Wager, S.%
\BCBL {}\ \BOthersPeriod {.}\end{APACrefauthors}%
\unskip\
\newblock
\APACrefYearMonthDay{2019}{}{}.
\newblock
{\BBOQ}\APACrefatitle {Generalized random forests} {Generalized random
  forests}.{\BBCQ}
\newblock
\APACjournalVolNumPages{The Annals of Statistics}{47}{2}{1148--1178}.
\PrintBackRefs{\CurrentBib}

\bibitem [\protect \citeauthoryear {%
Belloni%
, Chen%
, Chernozhukov%
\BCBL {}\ \BBA {} Hansen%
}{%
Belloni%
\ \protect \BOthers {.}}{%
{\protect \APACyear {2012}}%
}]{%
belloni2012sparse}
\APACinsertmetastar {%
belloni2012sparse}%
\begin{APACrefauthors}%
Belloni, A.%
, Chen, D.%
, Chernozhukov, V.%
\BCBL {}\ \BBA {} Hansen, C.%
\end{APACrefauthors}%
\unskip\
\newblock
\APACrefYearMonthDay{2012}{}{}.
\newblock
{\BBOQ}\APACrefatitle {Sparse models and methods for optimal instruments with
  an application to eminent domain} {Sparse models and methods for optimal
  instruments with an application to eminent domain}.{\BBCQ}
\newblock
\APACjournalVolNumPages{Econometrica}{80}{6}{2369--2429}.
\PrintBackRefs{\CurrentBib}

\bibitem [\protect \citeauthoryear {%
Belloni%
, Chernozhukov%
, Fern{\'a}ndez-Val%
\BCBL {}\ \BBA {} Hansen%
}{%
Belloni%
\ \protect \BOthers {.}}{%
{\protect \APACyear {2017}}%
}]{%
belloni2017program}
\APACinsertmetastar {%
belloni2017program}%
\begin{APACrefauthors}%
Belloni, A.%
, Chernozhukov, V.%
, Fern{\'a}ndez-Val, I.%
\BCBL {}\ \BBA {} Hansen, C.%
\end{APACrefauthors}%
\unskip\
\newblock
\APACrefYearMonthDay{2017}{}{}.
\newblock
{\BBOQ}\APACrefatitle {Program evaluation and causal inference with
  high-dimensional data} {Program evaluation and causal inference with
  high-dimensional data}.{\BBCQ}
\newblock
\APACjournalVolNumPages{Econometrica}{85}{1}{233--298}.
\PrintBackRefs{\CurrentBib}

\bibitem [\protect \citeauthoryear {%
Belloni%
, Chernozhukov%
\BCBL {}\ \BBA {} Hansen%
}{%
Belloni%
\ \protect \BOthers {.}}{%
{\protect \APACyear {2014}}%
}]{%
Belloni14restud}
\APACinsertmetastar {%
Belloni14restud}%
\begin{APACrefauthors}%
Belloni, A.%
, Chernozhukov, V.%
\BCBL {}\ \BBA {} Hansen, C.%
\end{APACrefauthors}%
\unskip\
\newblock
\APACrefYearMonthDay{2014}{}{}.
\newblock
{\BBOQ}\APACrefatitle {Inference on Treatment Effects after Selection among
  High-Dimensional Controls†} {Inference on treatment effects after selection
  among high-dimensional controls†}.{\BBCQ}
\newblock
\APACjournalVolNumPages{The Review of Economic Studies}{81}{2}{608-650}.
\PrintBackRefs{\CurrentBib}

\bibitem [\protect \citeauthoryear {%
Bitler%
, Gelbach%
\BCBL {}\ \BBA {} Hoynes%
}{%
Bitler%
\ \protect \BOthers {.}}{%
{\protect \APACyear {2006}}%
}]{%
bitler2006mean}
\APACinsertmetastar {%
bitler2006mean}%
\begin{APACrefauthors}%
Bitler, M\BPBI P.%
, Gelbach, J\BPBI B.%
\BCBL {}\ \BBA {} Hoynes, H\BPBI W.%
\end{APACrefauthors}%
\unskip\
\newblock
\APACrefYearMonthDay{2006}{}{}.
\newblock
{\BBOQ}\APACrefatitle {What mean impacts miss: Distributional effects of
  welfare reform experiments} {What mean impacts miss: Distributional effects
  of welfare reform experiments}.{\BBCQ}
\newblock
\APACjournalVolNumPages{American Economic Review}{96}{4}{988--1012}.
\PrintBackRefs{\CurrentBib}

\bibitem [\protect \citeauthoryear {%
Card%
}{%
Card%
}{%
{\protect \APACyear {1996}}%
}]{%
card1996effect}
\APACinsertmetastar {%
card1996effect}%
\begin{APACrefauthors}%
Card, D.%
\end{APACrefauthors}%
\unskip\
\newblock
\APACrefYearMonthDay{1996}{}{}.
\newblock
{\BBOQ}\APACrefatitle {The effect of unions on the structure of wages: A
  longitudinal analysis} {The effect of unions on the structure of wages: A
  longitudinal analysis}.{\BBCQ}
\newblock
\APACjournalVolNumPages{Econometrica: Journal of the Econometric
  Society}{}{}{957--979}.
\PrintBackRefs{\CurrentBib}

\bibitem [\protect \citeauthoryear {%
Chen%
, Genovese%
, Tibshirani%
, Wasserman%
\BCBL {}\ \protect \BOthers {.}}{%
Chen%
\ \protect \BOthers {.}}{%
{\protect \APACyear {2016}}%
}]{%
chen2016nonparametric}
\APACinsertmetastar {%
chen2016nonparametric}%
\begin{APACrefauthors}%
Chen, Y\BHBI C.%
, Genovese, C\BPBI R.%
, Tibshirani, R\BPBI J.%
, Wasserman, L.%
\BCBL {}\ \BOthersPeriod {.}\end{APACrefauthors}%
\unskip\
\newblock
\APACrefYearMonthDay{2016}{}{}.
\newblock
{\BBOQ}\APACrefatitle {Nonparametric modal regression} {Nonparametric modal
  regression}.{\BBCQ}
\newblock
\APACjournalVolNumPages{The Annals of Statistics}{44}{2}{489--514}.
\PrintBackRefs{\CurrentBib}

\bibitem [\protect \citeauthoryear {%
Chernozhukov%
\ \protect \BOthers {.}}{%
Chernozhukov%
\ \protect \BOthers {.}}{%
{\protect \APACyear {2018}}%
}]{%
chernozhukov2018double}
\APACinsertmetastar {%
chernozhukov2018double}%
\begin{APACrefauthors}%
Chernozhukov, V.%
, Chetverikov, D.%
, Demirer, M.%
, Duflo, E.%
, Hansen, C.%
, Newey, W.%
\BCBL {}\ \BBA {} Robins, J.%
\end{APACrefauthors}%
\unskip\
\newblock
\APACrefYearMonthDay{2018}{}{}.
\newblock
{\BBOQ}\APACrefatitle {Double/debiased machine learning for treatment and
  structural parameters} {Double/debiased machine learning for treatment and
  structural parameters}.{\BBCQ}
\newblock
\APACjournalVolNumPages{The Econometrics Journal}{21}{1}{C1--C68}.
\PrintBackRefs{\CurrentBib}

\bibitem [\protect \citeauthoryear {%
Chernozhukov%
, Escanciano%
, Ichimura%
\BCBL {}\ \BBA {} Newey%
}{%
Chernozhukov%
\ \protect \BOthers {.}}{%
{\protect \APACyear {2016}}%
}]{%
chernozhukov2016locally}
\APACinsertmetastar {%
chernozhukov2016locally}%
\begin{APACrefauthors}%
Chernozhukov, V.%
, Escanciano, J\BPBI C.%
, Ichimura, H.%
\BCBL {}\ \BBA {} Newey, W\BPBI K.%
\end{APACrefauthors}%
\unskip\
\newblock
\APACrefYearMonthDay{2016}{}{}.
\newblock
{\BBOQ}\APACrefatitle {Locally robust semiparametric estimation} {Locally
  robust semiparametric estimation}.{\BBCQ}
\newblock
\APACjournalVolNumPages{arXiv preprint arXiv:1608.00033}{}{}{}.
\PrintBackRefs{\CurrentBib}

\bibitem [\protect \citeauthoryear {%
Chernozhukov%
\ \BBA {} Hansen%
}{%
Chernozhukov%
\ \BBA {} Hansen%
}{%
{\protect \APACyear {2005}}%
}]{%
chernozhukov2005iv}
\APACinsertmetastar {%
chernozhukov2005iv}%
\begin{APACrefauthors}%
Chernozhukov, V.%
\BCBT {}\ \BBA {} Hansen, C.%
\end{APACrefauthors}%
\unskip\
\newblock
\APACrefYearMonthDay{2005}{}{}.
\newblock
{\BBOQ}\APACrefatitle {An IV model of quantile treatment effects} {An iv model
  of quantile treatment effects}.{\BBCQ}
\newblock
\APACjournalVolNumPages{Econometrica}{73}{1}{245--261}.
\PrintBackRefs{\CurrentBib}

\bibitem [\protect \citeauthoryear {%
Chernozhukov%
, Hansen%
\BCBL {}\ \BBA {} Spindler%
}{%
Chernozhukov%
\ \protect \BOthers {.}}{%
{\protect \APACyear {2015}}%
}]{%
chernozhukov2015valid}
\APACinsertmetastar {%
chernozhukov2015valid}%
\begin{APACrefauthors}%
Chernozhukov, V.%
, Hansen, C.%
\BCBL {}\ \BBA {} Spindler, M.%
\end{APACrefauthors}%
\unskip\
\newblock
\APACrefYearMonthDay{2015}{}{}.
\newblock
{\BBOQ}\APACrefatitle {Valid post-selection and post-regularization inference:
  An elementary, general approach} {Valid post-selection and
  post-regularization inference: An elementary, general approach}.{\BBCQ}
\newblock
\APACjournalVolNumPages{Annu. Rev. Econ.}{7}{1}{649--688}.
\PrintBackRefs{\CurrentBib}

\bibitem [\protect \citeauthoryear {%
DiNardo%
, Fortin%
\BCBL {}\ \BBA {} Lemieux%
}{%
DiNardo%
\ \protect \BOthers {.}}{%
{\protect \APACyear {1995}}%
}]{%
dinardo1995labor}
\APACinsertmetastar {%
dinardo1995labor}%
\begin{APACrefauthors}%
DiNardo, J.%
, Fortin, N\BPBI M.%
\BCBL {}\ \BBA {} Lemieux, T.%
\end{APACrefauthors}%
\unskip\
\newblock
\APACrefYearMonthDay{1995}{}{}.
\newblock
\APACrefbtitle {Labor market institutions and the distribution of wages,
  1973-1992: A semiparametric approach} {Labor market institutions and the
  distribution of wages, 1973-1992: A semiparametric approach}\
  \APACbVolEdTR{}{\BTR{}}.
\newblock
\APACaddressInstitution{}{National bureau of economic research}.
\PrintBackRefs{\CurrentBib}

\bibitem [\protect \citeauthoryear {%
Eddy%
\ \protect \BOthers {.}}{%
Eddy%
\ \protect \BOthers {.}}{%
{\protect \APACyear {1980}}%
}]{%
eddy1980optimum}
\APACinsertmetastar {%
eddy1980optimum}%
\begin{APACrefauthors}%
Eddy, W\BPBI F.%
\BCBT {}\ \BOthersPeriod {.}
\end{APACrefauthors}%
\unskip\
\newblock
\APACrefYearMonthDay{1980}{}{}.
\newblock
{\BBOQ}\APACrefatitle {Optimum kernel estimators of the mode} {Optimum kernel
  estimators of the mode}.{\BBCQ}
\newblock
\APACjournalVolNumPages{The Annals of Statistics}{8}{4}{870--882}.
\PrintBackRefs{\CurrentBib}

\bibitem [\protect \citeauthoryear {%
Firpo%
}{%
Firpo%
}{%
{\protect \APACyear {2007}}%
}]{%
firpo2007efficient}
\APACinsertmetastar {%
firpo2007efficient}%
\begin{APACrefauthors}%
Firpo, S.%
\end{APACrefauthors}%
\unskip\
\newblock
\APACrefYearMonthDay{2007}{}{}.
\newblock
{\BBOQ}\APACrefatitle {Efficient semiparametric estimation of quantile
  treatment effects} {Efficient semiparametric estimation of quantile treatment
  effects}.{\BBCQ}
\newblock
\APACjournalVolNumPages{Econometrica}{75}{1}{259--276}.
\PrintBackRefs{\CurrentBib}

\bibitem [\protect \citeauthoryear {%
Freeman%
}{%
Freeman%
}{%
{\protect \APACyear {1980}}%
}]{%
freeman1980unionism}
\APACinsertmetastar {%
freeman1980unionism}%
\begin{APACrefauthors}%
Freeman, R\BPBI B.%
\end{APACrefauthors}%
\unskip\
\newblock
\APACrefYearMonthDay{1980}{}{}.
\newblock
{\BBOQ}\APACrefatitle {Unionism and the Dispersion of Wages} {Unionism and the
  dispersion of wages}.{\BBCQ}
\newblock
\APACjournalVolNumPages{ILR Review}{34}{1}{3--23}.
\PrintBackRefs{\CurrentBib}

\bibitem [\protect \citeauthoryear {%
Hahn%
}{%
Hahn%
}{%
{\protect \APACyear {1998}}%
}]{%
hahn1998role}
\APACinsertmetastar {%
hahn1998role}%
\begin{APACrefauthors}%
Hahn, J.%
\end{APACrefauthors}%
\unskip\
\newblock
\APACrefYearMonthDay{1998}{}{}.
\newblock
{\BBOQ}\APACrefatitle {On the role of the propensity score in efficient
  semiparametric estimation of average treatment effects} {On the role of the
  propensity score in efficient semiparametric estimation of average treatment
  effects}.{\BBCQ}
\newblock
\APACjournalVolNumPages{Econometrica}{}{}{315--331}.
\PrintBackRefs{\CurrentBib}

\bibitem [\protect \citeauthoryear {%
Hansen%
}{%
Hansen%
}{%
{\protect \APACyear {2008}}%
}]{%
hansen2008uniform}
\APACinsertmetastar {%
hansen2008uniform}%
\begin{APACrefauthors}%
Hansen, B\BPBI E.%
\end{APACrefauthors}%
\unskip\
\newblock
\APACrefYearMonthDay{2008}{}{}.
\newblock
{\BBOQ}\APACrefatitle {Uniform convergence rates for kernel estimation with
  dependent data} {Uniform convergence rates for kernel estimation with
  dependent data}.{\BBCQ}
\newblock
\APACjournalVolNumPages{Econometric Theory}{24}{3}{726--748}.
\PrintBackRefs{\CurrentBib}

\bibitem [\protect \citeauthoryear {%
Heckman%
, Ichimura%
\BCBL {}\ \BBA {} Todd%
}{%
Heckman%
\ \protect \BOthers {.}}{%
{\protect \APACyear {1997}}%
}]{%
heckman1997matching}
\APACinsertmetastar {%
heckman1997matching}%
\begin{APACrefauthors}%
Heckman, J.%
, Ichimura, H.%
\BCBL {}\ \BBA {} Todd, P\BPBI E.%
\end{APACrefauthors}%
\unskip\
\newblock
\APACrefYearMonthDay{1997}{}{}.
\newblock
{\BBOQ}\APACrefatitle {Matching as an econometric evaluation estimator:
  Evidence from evaluating a job training programme} {Matching as an
  econometric evaluation estimator: Evidence from evaluating a job training
  programme}.{\BBCQ}
\newblock
\APACjournalVolNumPages{The review of economic studies}{64}{4}{605--654}.
\PrintBackRefs{\CurrentBib}

\bibitem [\protect \citeauthoryear {%
Heckman%
\ \BBA {} Robb%
}{%
Heckman%
\ \BBA {} Robb%
}{%
{\protect \APACyear {1985}}%
}]{%
heckman1985alternative}
\APACinsertmetastar {%
heckman1985alternative}%
\begin{APACrefauthors}%
Heckman, J.%
\BCBT {}\ \BBA {} Robb, R.%
\end{APACrefauthors}%
\unskip\
\newblock
\APACrefYearMonthDay{1985}{}{}.
\newblock
{\BBOQ}\APACrefatitle {Alternative methods for evaluating the impact of
  interventions: An overview} {Alternative methods for evaluating the impact of
  interventions: An overview}.{\BBCQ}
\newblock
\APACjournalVolNumPages{Journal of econometrics}{30}{1-2}{239--267}.
\PrintBackRefs{\CurrentBib}

\bibitem [\protect \citeauthoryear {%
Hirano%
, Imbens%
\BCBL {}\ \BBA {} Ridder%
}{%
Hirano%
\ \protect \BOthers {.}}{%
{\protect \APACyear {2003}}%
}]{%
hirano2003efficient}
\APACinsertmetastar {%
hirano2003efficient}%
\begin{APACrefauthors}%
Hirano, K.%
, Imbens, G\BPBI W.%
\BCBL {}\ \BBA {} Ridder, G.%
\end{APACrefauthors}%
\unskip\
\newblock
\APACrefYearMonthDay{2003}{}{}.
\newblock
{\BBOQ}\APACrefatitle {Efficient estimation of average treatment effects using
  the estimated propensity score} {Efficient estimation of average treatment
  effects using the estimated propensity score}.{\BBCQ}
\newblock
\APACjournalVolNumPages{Econometrica}{71}{4}{1161--1189}.
\PrintBackRefs{\CurrentBib}

\bibitem [\protect \citeauthoryear {%
Lee%
}{%
Lee%
}{%
{\protect \APACyear {1989}}%
}]{%
lee1989mode}
\APACinsertmetastar {%
lee1989mode}%
\begin{APACrefauthors}%
Lee, M\BHBI J.%
\end{APACrefauthors}%
\unskip\
\newblock
\APACrefYearMonthDay{1989}{}{}.
\newblock
{\BBOQ}\APACrefatitle {Mode regression} {Mode regression}.{\BBCQ}
\newblock
\APACjournalVolNumPages{Journal of Econometrics}{42}{3}{337--349}.
\PrintBackRefs{\CurrentBib}

\bibitem [\protect \citeauthoryear {%
Newey%
\ \BBA {} McFadden%
}{%
Newey%
\ \BBA {} McFadden%
}{%
{\protect \APACyear {1994}}%
}]{%
newey1994large}
\APACinsertmetastar {%
newey1994large}%
\begin{APACrefauthors}%
Newey, W\BPBI K.%
\BCBT {}\ \BBA {} McFadden, D.%
\end{APACrefauthors}%
\unskip\
\newblock
\APACrefYearMonthDay{1994}{}{}.
\newblock
{\BBOQ}\APACrefatitle {Large sample estimation and hypothesis testing} {Large
  sample estimation and hypothesis testing}.{\BBCQ}
\newblock
\APACjournalVolNumPages{Handbook of econometrics}{4}{}{2111--2245}.
\PrintBackRefs{\CurrentBib}

\bibitem [\protect \citeauthoryear {%
Parzen%
}{%
Parzen%
}{%
{\protect \APACyear {1962}}%
}]{%
parzen1962estimation}
\APACinsertmetastar {%
parzen1962estimation}%
\begin{APACrefauthors}%
Parzen, E.%
\end{APACrefauthors}%
\unskip\
\newblock
\APACrefYearMonthDay{1962}{}{}.
\newblock
{\BBOQ}\APACrefatitle {On estimation of a probability density function and
  mode} {On estimation of a probability density function and mode}.{\BBCQ}
\newblock
\APACjournalVolNumPages{The annals of mathematical
  statistics}{33}{3}{1065--1076}.
\PrintBackRefs{\CurrentBib}

\bibitem [\protect \citeauthoryear {%
Robins%
\ \BBA {} Rotnitzky%
}{%
Robins%
\ \BBA {} Rotnitzky%
}{%
{\protect \APACyear {1995}}%
}]{%
robins1995semiparametric}
\APACinsertmetastar {%
robins1995semiparametric}%
\begin{APACrefauthors}%
Robins, J\BPBI M.%
\BCBT {}\ \BBA {} Rotnitzky, A.%
\end{APACrefauthors}%
\unskip\
\newblock
\APACrefYearMonthDay{1995}{}{}.
\newblock
{\BBOQ}\APACrefatitle {Semiparametric efficiency in multivariate regression
  models with missing data} {Semiparametric efficiency in multivariate
  regression models with missing data}.{\BBCQ}
\newblock
\APACjournalVolNumPages{Journal of the American Statistical
  Association}{90}{429}{122--129}.
\PrintBackRefs{\CurrentBib}

\bibitem [\protect \citeauthoryear {%
Rosenbaum%
\ \BBA {} Rubin%
}{%
Rosenbaum%
\ \BBA {} Rubin%
}{%
{\protect \APACyear {1983}}%
}]{%
rosenbaum1983central}
\APACinsertmetastar {%
rosenbaum1983central}%
\begin{APACrefauthors}%
Rosenbaum, P\BPBI R.%
\BCBT {}\ \BBA {} Rubin, D\BPBI B.%
\end{APACrefauthors}%
\unskip\
\newblock
\APACrefYearMonthDay{1983}{}{}.
\newblock
{\BBOQ}\APACrefatitle {The central role of the propensity score in
  observational studies for causal effects} {The central role of the propensity
  score in observational studies for causal effects}.{\BBCQ}
\newblock
\APACjournalVolNumPages{Biometrika}{70}{1}{41--55}.
\PrintBackRefs{\CurrentBib}

\bibitem [\protect \citeauthoryear {%
Tauchen%
}{%
Tauchen%
}{%
{\protect \APACyear {1985}}%
}]{%
tauchen1985diagnostic}
\APACinsertmetastar {%
tauchen1985diagnostic}%
\begin{APACrefauthors}%
Tauchen, G.%
\end{APACrefauthors}%
\unskip\
\newblock
\APACrefYearMonthDay{1985}{}{}.
\newblock
{\BBOQ}\APACrefatitle {Diagnostic testing and evaluation of maximum likelihood
  models} {Diagnostic testing and evaluation of maximum likelihood
  models}.{\BBCQ}
\newblock
\APACjournalVolNumPages{Journal of Econometrics}{30}{1-2}{415--443}.
\PrintBackRefs{\CurrentBib}

\bibitem [\protect \citeauthoryear {%
Van~der Vaart%
}{%
Van~der Vaart%
}{%
{\protect \APACyear {2000}}%
}]{%
van2000asymptotic}
\APACinsertmetastar {%
van2000asymptotic}%
\begin{APACrefauthors}%
Van~der Vaart, A\BPBI W.%
\end{APACrefauthors}%
\unskip\
\newblock
\APACrefYear{2000}.
\newblock
\APACrefbtitle {Asymptotic statistics} {Asymptotic statistics}\ (\BVOL~3).
\newblock
\APACaddressPublisher{}{Cambridge university press}.
\PrintBackRefs{\CurrentBib}

\bibitem [\protect \citeauthoryear {%
Yao%
\ \BBA {} Li%
}{%
Yao%
\ \BBA {} Li%
}{%
{\protect \APACyear {2014}}%
}]{%
yao2014new}
\APACinsertmetastar {%
yao2014new}%
\begin{APACrefauthors}%
Yao, W.%
\BCBT {}\ \BBA {} Li, L.%
\end{APACrefauthors}%
\unskip\
\newblock
\APACrefYearMonthDay{2014}{}{}.
\newblock
{\BBOQ}\APACrefatitle {A new regression model: modal linear regression} {A new
  regression model: modal linear regression}.{\BBCQ}
\newblock
\APACjournalVolNumPages{Scandinavian Journal of Statistics}{41}{3}{656--671}.
\PrintBackRefs{\CurrentBib}

\end{thebibliography}

\section{Appendix}
\textit{Proof of Theorem 1:} We only present the proof of the first claim, $\hat{\theta}_{1}\overset{p}{\to} \theta_{1}^{*}$, since the second claim follows from the same arguments. The proof proceeds in two steps. In Step 1, we show the uniform law of large number holds
\[\sup_{y}\mid \hat{f}_{Y_{1}}\left(y\right)-f_{Y_{1}}\left(y\right)\mid=o_{p}\left(1\right).
\]
In Step 2, we establish the consistency $\hat{\theta}_{1}\overset{p}{\to} \theta_{1}^{*}$ using the same argument of Theorem 5.7 in \citet*{van2000asymptotic}.

\textit{Step 1.} 
Notice that we have the decomposition
\begin{align*}
\hat{f}_{Y_{1}}\left(y\right)-f_{Y_{1}}\left(y\right)& = \frac{1}{n}\sum_{i=1}^{n}\hat{f}_{Y\mid D=1,X}\left(y\mid X_{i}\right)-E\left[ f_{Y\mid D=1,X}\left(y\mid X\right)\right]\\
 & = \underbrace{\frac{1}{n}\sum_{i=1}^{n}\left(\hat{f}_{Y\mid D=1,X}\left(y\mid X_{i}\right)-f_{Y\mid D=1,X}\left(y\mid X_{i}\right)\right)}_{A\left(y\right)}\\
  & + \underbrace{\frac{1}{n}\sum_{i=1}^{n}f_{Y\mid D=1,X}\left(y\mid X_{i}\right)-E\left[ f_{Y\mid D=1,X}\left(y\mid X\right)\right]}_{B\left(y\right)}.
\end{align*}
Hence, 
\[
\sup_{y}\mid \hat{f}_{Y_{1}}\left(y\right)-f_{Y_{1}}\left(y\right) \mid\leq \sup_{y}\mid A\left(y\right)\mid +\sup_{y}\mid B\left(y\right)\mid
\]
By Theorem 6 in \citet*{hansen2008uniform} (uniform rates of convergence of kernel estimators), the first term $\sup_{y}\mid A\left(y\right)\mid$ is bounded by
\begin{align*}
\sup_{y}\abs{A\left(y\right)} & \leq \sup_{y} \frac{1}{n}\sum_{i=1}^{n}\abs{\hat{f}_{Y\mid D=1,X}\left(y\mid X_{i}\right)-f_{Y\mid D=1,X}\left(y\mid X_{i}\right)}\\
& \leq \sup_{y}\sup_{x} \abs{\hat{f}_{Y\mid D=1,X}\left(y\mid x\right)-f_{Y\mid D=1,X}\left(y\mid x\right)} \\
& \leq \sup_{x,y}\abs{\hat{f}_{Y\mid D=1,X}\left(y\mid x\right)-f_{Y\mid D=1,X}\left(y\mid x\right)}\\
& = O_{p}\left(\sqrt{\frac{\ln{n}}{nh^{d+1}}}+h^2\right)\\
& = o_{p}\left(1\right).
\end{align*} 
On the other hand, by Lemma 1 of \citet*{tauchen1985diagnostic} (uniform law of large numbers), we have

\[
\sup_{y\in \mathcal{Y}}\abs{B\left(y\right)}=\sup_{y\in \mathcal{Y}}\abs{\frac{1}{n}\sum_{i=1}^{n}f_{Y\mid D=1,X}\left(y\mid X_{i}\right)-E\left[ f_{Y\mid D=1,X}\left(y\mid X\right)\right]}\overset{p}{\to}0.
\]
Combining the results of $\sup_{y}\mid A\left(y\right)\mid$ and $\sup_{y}\mid B\left(y\right)\mid$ gives 

\[
\sup_{y}\mid \hat{f}_{Y_{1}}\left(y\right)-f_{Y_{1}}\left(y\right) \mid=o_{p}\left(1\right). 
\]

\textit{Step 2.} The definition of $\hat{\theta}_{1}$ implies that $\hat{f}_{Y_{1}}\left(\hat{\theta}_{1}\right)\geq \hat{f}_{Y_{1}}\left(\theta_{1}^{*}\right)$. Therefore, we have

\begin{align*}
f_{Y_{1}}\left(\theta_{1}^{*}\right)-f_{Y_{1}}\left(\hat{\theta}_{1}\right) & =  f_{Y_{1}}\left(\theta_{1}^{*}\right)-\hat{f}_{Y_{1}}\left(\theta_{1}^{*}\right)+\hat{f}_{Y_{1}}\left(\theta_{1}^{*}\right)-f_{Y_{1}}\left(\hat{\theta}_{1}\right)\\
 & \leq  f_{Y_{1}}\left(\theta_{1}^{*}\right)-\hat{f}_{Y_{1}}\left(\theta_{1}^{*}\right)+\hat{f}_{Y_{1}}\left(\hat{\theta}_{1}\right)-f_{Y_{1}}\left(\hat{\theta}_{1}\right)\\
 & \leq 2 \sup_{y}\mid \hat{f}_{Y_{1}}\left(y\right)-f_{Y_{1}}\left(y\right)\mid.
\end{align*}
By Step 1, we have that for any $\delta>0$,
\[
P\left(f_{Y_{1}}\left(\theta_{1}^{*}\right)-f_{Y_{1}}\left(\hat{\theta}_{1}\right)>\delta\right)\leq P\left(\sup_{y}\mid \hat{f}_{Y_{1}}\left(y\right)-f_{Y_{1}}\left(y\right)\mid>\delta/2\right)\to  0. 
\]
Further, Assumption 1 implies that for any $\varepsilon>0$, there exists $\delta>0$ such that
\[
\sup_{y:\mid y-\theta_{1}^{*}\mid>\varepsilon}f_{Y_{1}}\left(y\right)<f_{Y_{1}}\left(\theta_{1}^{*}\right)-\delta.
\]
Then the following inequality holds
\begin{align*}
P\left(\mid \hat{\theta}_{1}-\theta_{1}^{*}\mid > \varepsilon\right)& \leq P\left(f_{Y_{1}}\left(\hat{\theta}_{1}\right)<f_{Y_{1}}\left(\theta_{1}^{*}\right)-\delta\right)\\
 & \leq P\left(f_{Y_{1}}\left(\theta_{1}^{*}\right)-f_{Y_{1}}\left(\hat{\theta}_{1}\right)>\delta\right)\to  0.
\end{align*}
Thus, we prove the consistency $\hat{\theta}_{1}\overset{p}{\to}\theta_{1}^{*}$.
\vskip 1cm

\textit{Proof of Theorem 2:} Here we focus on the result for $\hat{\theta}_{1}$ only. Notice that the first-order condition for $\hat{\theta}_{1}$ gives

\begin{align*}
0 & =\hat{f}_{Y_{1}}^{\left(1\right)}\left(\hat{\theta}_{1}\right)=\frac{1}{n}\sum_{i=1}^{n}\hat{f}_{Y\mid X,D=1}^{\left(1\right)}\left(\hat{\theta}_{1}\mid X_{i}\right)\\
 & = \frac{1}{n}\sum_{i=1}^{n}\hat{f}_{Y\mid X,D=1}^{\left(1\right)}\left(\theta_{1}^{*}\mid X_{i}\right)+\frac{1}{n}\sum_{i=1}^{n}\hat{f}_{Y\mid X,D=1}^{\left(2\right)}\left(\tilde{\theta}_{1}\mid X_{i}\right)\left(\hat{\theta}_{1}-\theta_{1}^{*}\right),
\end{align*}
where $\tilde{\theta}_{1}\in \left(\hat{\theta}_{1},\theta_{1}^{*}\right)$. Then we have
\[
\sqrt{nh^{3}}\left(\hat{\theta}_{1}-\theta_{1}^{*}\right)=-\left[\frac{1}{n}\sum_{i=1}^{n}\hat{f}_{Y\mid X,D=1}^{\left(2\right)}\left(\tilde{\theta}_{1}\mid X_{i}\right)\right]^{-1}\left(\frac{\sqrt{nh^{3}}}{n}\sum_{i=1}^{n}\hat{f}_{Y\mid X,D=1}^{\left(1\right)}\left(\theta_{1}^{*}\mid X_{i}\right)\right).
\]
The proof proceeds in six steps. In Step 1, we show that the first term of r.h.s converges to $M_{1}=E\left[f_{Y\mid X,D=1}^{\left(2\right)}\left(\theta_{1}^{*}\mid X\right)\right]$ in probability. In Step 2-5, we show the asymptotic normality of the second term. Then, by Slutsky's theorem, we can show the asymptotic normality for $\hat{\theta}_{1}$. In Step 6, we show the asymptotic normality for $\hat{\Delta}$.

For convenience, we define $\gamma_{10}\left(x\right)\equiv f_{Y,X\mid D=1}^{\left(1\right)}\left(\theta_{1}^{*},x\right)$, $\gamma_{20}\left(x\right)\equiv f_{X\mid D=1}\left(x\right)$, and 
\begin{align*}
&\hat{\gamma}_{1}\left(x\right)\equiv \frac{1}{n}\sum_{j=1}^{n}\frac{D_{j}K_{h}^{\left(1\right)}\left(\theta_{1}^{*}-Y_{j}\right)K_{h}\left(x-X_{j}\right)}{P\left(D=1\right)}\\
&\hat{\gamma}_{2}\left(x\right)\equiv \frac{1}{n}\sum_{j=1}^{n}\frac{D_{j}K_{h}\left(x-X_{j}\right)}{P\left(D=1\right)}.
\end{align*}
In these notations, we can express $\hat{f}_{Y\mid X,D=1}^{\left(1\right)}\left(\theta_{1}^{*}\mid x\right)$ and $f_{Y\mid X,D=1}^{\left(1\right)}\left(\theta_{1}^{*}\mid x\right)$ as $\hat{\gamma}_{1}\left(x\right)/\hat{\gamma}_{2}\left(x\right)$ and $\gamma_{10}\left(x\right)/\gamma_{20}\left(x\right)$, respectively. Also, let $\gamma_{0}=\left(\gamma_{10},\gamma_{20}\right)'$ and $\hat{\gamma}=\left(\hat{\gamma}_{1},\hat{\gamma}_{2}\right)'$.

\textit{Step 1.}
In this step, we show that $n^{-1}\sum_{i=1}^{n}\hat{f}_{Y\mid X, D=1}^{\left(2\right)}\left(\tilde{\theta}_{1}\mid X_{i}\right)\overset{p}{\to}E\left[f_{Y\mid X,D=1}^{\left(2\right)}\left(\theta_{1}^{*}\mid X\right)\right]$.
Notice that
\[
\frac{1}{n}\sum_{i=1}^{n}\hat{f}_{Y\mid X, D=1}^{\left(2\right)}\left(\tilde{\theta}_{1}\mid X_{i}\right)=\frac{1}{n}\sum_{i=1}^{n}f_{Y\mid X, D=1}^{\left(2\right)}\left(\theta_{1}^{*}\mid X_{i}\right)+A_{1}+A_{2}
\]
where 
\[
A_{1}=\frac{1}{n}\sum_{i=1}^{n}\hat{f}_{Y\mid X, D=1}^{\left(2\right)}\left(\tilde{\theta}_{1}\mid X_{i}\right)-f_{Y\mid X, D=1}^{\left(2\right)}\left(\tilde{\theta}_{1}\mid X_{i}\right)
\]
and
\[
A_{2}=\frac{1}{n}\sum_{i=1}^{n}f_{Y\mid X, D=1}^{\left(2\right)}\left(\tilde{\theta}_{1}\mid X_{i}\right)-f_{Y\mid X, D=1}^{\left(2\right)}\left(\theta_{1}^{*}\mid X_{i}\right).
\]
Since $\frac{1}{n}\sum_{i=1}^{n}f_{Y\mid X, D=1}^{\left(2\right)}\left(\theta_{1}^{*}\mid X_{i}\right)\overset{p}{\to}E\left[f_{Y\mid X,D=1}^{\left(2\right)}\left(\theta_{1}^{*}\mid X\right)\right]$ by the law of large numbers, we only have to show that $A_{1}=o_{p}\left(1\right)$ and $A_{2}=o_{p}\left(1\right)$. Note that 
\begin{align*}
\abs{A_{1}}& \leq \frac{1}{n}\sum_{i=1}^{n}\abs{\hat{f}_{Y\mid X, D=1}^{\left(2\right)}\left(\tilde{\theta}_{1}\mid X_{i}\right)-f_{Y\mid X, D=1}^{\left(2\right)}\left(\tilde{\theta}_{1}\mid X_{i}\right)}\\
& \leq \sup_{y,x}\abs{\hat{f}_{Y\mid X, D=1}^{\left(2\right)}\left(y\mid x\right)-f_{Y\mid X, D=1}^{\left(2\right)}\left(y\mid x\right)}\\
& = O_{p}\left(\sqrt{\frac{\ln{n}}{nh^{d+5}}}+h^{2}\right)\\
& = o_{p}\left(1\right),
\end{align*}
where the first equality follows from the uniform rates of convergence of kernel estimators \citep*{hansen2008uniform}. For $A_{2}$, we use the argument in Lemma 4.3 of \citet*{newey1994large}. By consistency of $\hat{\theta}_{1}$, and thus $\tilde{\theta}_{1}$, there is $\delta_{n}\to 0$ such that $\norm{\tilde{\theta}_{1}-\theta_{1}^{*}}\leq \delta_{n}$ with probability approaching to one. Define 

\[\Delta_{n}\left(X_{i}\right)=\sup_{\norm{y-\theta_{1}^{*}}\leq \delta_{n}}\norm{f_{Y\mid X, D=1}^{\left(2\right)}\left(y\mid X_{i}\right)-f_{Y\mid X, D=1}^{\left(2\right)}\left(\theta_{1}^{*}\mid X_{i}\right)}.
\] 
By the continuity of $f_{Y\mid X, D=1}^{\left(2\right)}\left(y\mid X_{i}\right)$ at $\theta_{1}^{*}$, $\Delta_{n}\left(X_{i}\right)\overset{p}{\to} 0$. Hence, by the dominated convergence theorem, we have $E\left[\Delta_{n}\left(X_{i}\right)\right]\to 0
$. Then, by Markov's inequality, 
\[
P\left(\frac{1}{n}\sum_{i=1}^{n}\Delta_{n}\left(X_{i}\right)> \epsilon\right)\leq E\left[\Delta_{n}\left(X_{i}\right)\right]/\epsilon\to 0
.\] 
Therefore, we have

\[\abs{A_{2}}\leq \frac{1}{n}\sum_{i=1}^{n}\Delta_{n}\left(X_{i}\right)+o_{p}\left(1\right)=o_{p}\left(1\right).
\] 

\textit{Step 2.}
In this step, we show 
\[
\frac{\sqrt{nh^{3}}}{n}\sum_{i=1}^{n}\hat{f}_{Y\mid X,D=1}^{\left(1\right)}\left(\theta_{1}^{*}\mid X_{i}\right)  = \frac{\sqrt{nh^{3}}}{n}\sum_{i=1}^{n}f_{Y\mid X,D=1}^{\left(1\right)}\left(\theta_{1}^{*}\mid X_{i}\right)+\frac{\sqrt{nh^{3}}}{n}\sum_{i=1}^{n} G\left(Z_{i},\hat{\gamma}-
\gamma_{0}\right)+o_{p}\left(1\right),
\]
where $G\left(z,\gamma\right)=\gamma_{20}\left(x\right)^{-1}\left[1,-\frac{\gamma_{10}\left(x\right)}{\gamma_{20}\left(x\right)}\right]\gamma\left(x\right)$ and $z=\left(y,x,d\right)$ denotes data observation. 
To do this, it suffices to show 
\[
\frac{\sqrt{nh^{3}}}{n}\sum_{i=1}^{n}\left[\hat{f}_{Y\mid X,D=1}^{\left(1\right)}\left(\theta_{1}^{*}\mid X_{i}\right)-f_{Y\mid X,D=1}^{\left(1\right)}\left(\theta_{1}^{*}\mid X_{i}\right)- G\left(Z_{i},\hat{\gamma}-
\gamma_{0}\right)\right]=o_{p}\left(1\right).
\]

Using the notation of $\gamma$, we have
\begin{align*}
\hat{f}_{Y\mid X,D=1}^{\left(1\right)}\left(\theta_{1}^{*}\mid x\right)-f_{Y\mid X,D=1}^{\left(1\right)}\left(\theta_{1}^{*}\mid x\right) & = \frac{\hat{\gamma}_{1}\left(x\right)}{\hat{\gamma}_{2}\left(x\right)}-\frac{\gamma_{10}\left(x\right)}{\gamma_{20}\left(x\right)}.
\end{align*}
The following argument follows from \citet*{newey1994large}. Consider the algebra relation $\tilde{a}/\tilde{b}-a/b=b^{-1}\left[1-\tilde{b}^{-1}\left(\tilde{b}-b\right)\right]\left[\tilde{a}-a-\left(a/b\right)\left(\tilde{b}-b\right)\right]$. The linear part of the r.h.s is $b^{-1}\left[\tilde{a}-a-\left(a/b\right)\left(\tilde{b}-b\right)\right]$, and the remaining term is of higher order. By letting $a=\gamma_{10}$, $\tilde{a}=\hat{\gamma}_{1}$, $b=\gamma_{20}$, and $\tilde{b}=\hat{\gamma}_{2}$, this linear term corresponds to the linear functional $G\left(Z_{i},\hat{\gamma}-
\gamma_{0}\right)$.  The remaining higher-order term will satisfy 

\begin{align*}
& \mid \frac{\gamma_{1}\left(x\right)}{\gamma_{2}\left(x\right)}-\frac{\gamma_{10}\left(x\right)}{\gamma_{20}\left(x\right)}-G\left(z,\gamma-
\gamma_{0}\right)\mid  \\
& \leq \mid \gamma_{2}\left(x\right)\mid^{-1}\gamma_{20}\left(x\right)^{-1}\left[1+\frac{\gamma_{10}\left(x\right)}{\gamma_{20}\left(x\right)}\right]\left[\left(\gamma_{1}\left(x\right)-\gamma_{10}\left(x\right)\right)^{2}+\left(\gamma_{2}\left(x\right)-\gamma_{20}\left(x\right)\right)^{2}\right]\\
& \leq C\sup_{x\in \mathcal{X}}\left\lVert \gamma\left(x\right)-
\gamma_{0}\left(x\right)\right\lVert^{2}
\end{align*}
for some constant $C$ if $\gamma_{2}$ and $\gamma_{20}$ are bounded away from zero. Hence Lemma 1 holds if $\sqrt{nh^{3}}\sup_{x\in \mathcal{X}}\left\lVert \hat{\gamma}\left(x\right)-
\gamma_{0}\left(x\right)\right\lVert^{2}\overset{p}\to0$. By the uniform rates of convergence of  kernel estimators \citep*{hansen2008uniform}, we have
\begin{align*}
\sup_{x\in \mathcal{X}}\left\lVert \hat{\gamma}\left(x\right)-
\gamma_{0}\left(x\right)\right\lVert^{2}& =\sup_{x\in \mathcal{X}}\left(\left( \hat{\gamma}_{1}\left(x\right)-
\gamma_{10}\left(x\right)\right)^{2}+\left( \hat{\gamma}_{1}\left(x\right)-
\gamma_{10}\left(x\right)\right)^{2}\right)\\
& \leq \sup_{x\in \mathcal{X}} \left(\hat{\gamma}_{1}\left(x\right)-
\gamma_{10}\left(x\right)\right)^{2}+ \sup_{x\in \mathcal{X}} \left(\hat{\gamma}_{2}\left(x\right)-
\gamma_{20}\left(x\right)\right)^{2}\\
& = O_{p}\left[\left(\ln{n} \right)\left(nh^{d+3} \right)^{-1}+h^{4}\right]+O_{p}\left[\left(\ln{n} \right)\left(nh^{d} \right)^{-1}+h^{4}\right]\\
& = O_{p}\left[\left(\ln{n} \right)\left(nh^{d+3} \right)^{-1}+h^{4}\right].
\end{align*}
The rates of $h$ and $n$ imply that $\sqrt{nh^{3}}\sup_{x\in \mathcal{X}}\left\lVert \hat{\gamma}\left(x\right)-
\gamma_{0}\left(x\right)\right\lVert^{2}\overset{p}\to 0$.

\textit{Step 3.}
In this step, we show 
\begin{align*}
\frac{\sqrt{nh^{3}}}{n}\sum_{i=1}^{n}\hat{f}_{Y\mid X,D=1}^{\left(1\right)}\left(\theta_{1}^{*}\mid X_{i}\right)& = \frac{\sqrt{nh^{3}}}{n}\sum_{i=1}^{n}f_{Y\mid X,D=1}^{\left(1\right)}\left(\theta_{1}^{*}\mid X_{i}\right)+\sqrt{nh^{3}}\int G\left(z,\hat{\gamma}-
\gamma_{0}\right)dF_{0}\left(z\right)\\
& +o_{p}\left(1\right),
\end{align*}
where $F_{0}$ is the c.d.f. of $z$. To do this, it suffices to show that 
\[
\sqrt{nh^{3}}\left\{\frac{1}{n}\sum_{i=1}^{n}G\left(Z_{i}, \hat{\gamma}-\gamma_{0}\right)-\int G\left(z, \hat{\gamma}-\gamma_{0}\right)dF_{0}\left(z\right)\right\}=o_{p}\left(1\right).
\]

Let $\bar{\gamma}\equiv E\left[\hat{\gamma}\right]$ and by the linearity of $G\left(z, \gamma\right)$, we have the decomposition $$G\left(z, \hat{\gamma}-\gamma_{0}\right)=G\left(z, \hat{\gamma}-\bar{\gamma}\right)+G\left(z, \bar{\gamma}-\gamma_{0}\right).$$
Therefore we just need to show that
\begin{align*}
\sqrt{nh^{3}}\left\{\frac{1}{n}\sum_{i=1}^{n}G\left(Z_{i}, \hat{\gamma}-\bar{\gamma}\right)-\int G\left(z, \hat{\gamma}-\bar{\gamma}\right)dF_{0}\left(z\right)\right\}=o_{p}\left(1\right)
\end{align*}
and 
\[
\sqrt{nh^{3}}\left\{\frac{1}{n}\sum_{i=1}^{n}G\left(Z_{i}, \bar{\gamma}-\gamma_{0}\right)-\int G\left(z, \bar{\gamma}-\gamma_{0}\right)dF_{0}\left(z\right)\right\}=o_{p}\left(1\right).
\]
The second condition holds by the central limit theorem since
\[
\sqrt{nh^{3}}\left\{\frac{1}{n}\sum_{i=1}^{n}G\left(Z_{i}, \bar{\gamma}-\gamma_{0}\right)-\int G\left(z, \bar{\gamma}-\gamma_{0}\right)dF_{0}\left(z\right)\right\}=\sqrt{nh^{3}}O_{p}\left(n^{-1/2}\right)=o_{p}\left(1\right).
\]
It remains to show the first condition. We follow the arguments in \citet*{newey1994large}. 
Define $q_{j}\equiv \left(\frac{D_{j}K_{h}^{\left(1\right)}\left(\theta_{1}^{*}-Y_{j}\right)}{P\left(D=1\right)}, \frac{D_{j}}{P\left(D=1\right)}\right)'$, we can rewrite
\begin{align*}
\hat{\gamma}\left(x\right)=\begin{bmatrix}
\hat{\gamma}_{1}\left(x\right)\\
\hat{\gamma}_{2}\left(x\right)
\end{bmatrix}=\frac{1}{n}\sum_{j=1}^{n}q_{j}K_{h}\left(x-X_{j}\right).
\end{align*}
We also define 
\begin{align*}
m\left(Z_{i}, Z_{j}\right)& =G\left[Z_{i}, q_{j}K_{h}\left(\cdot - X_{j}\right)\right]\\
m_{1}\left(z\right)& =\int m\left(z, \tilde{z}\right)dF_{0}\left(\tilde{z}\right)=G\left(z,\bar{\gamma}\right)\\
m_{2}\left(z\right)& =\int m\left(\tilde{z}, z\right)dF_{0}\left(\tilde{z}\right)=\int G\left[\tilde{z}, qK_{h}\left(\cdot - X\right)\right]dF_{0}\left(\tilde{z}\right).
\end{align*}
Then the l.h.s. of the first condition equals

\begin{align*}
& \sqrt{nh^{3}}\left\{\frac{1}{n}\sum_{i=1}^{n}G\left(Z_{i}, \hat{\gamma}-\bar{\gamma}\right)-\int G\left(z, \hat{\gamma}-\bar{\gamma}\right)dF_{0}\left(z\right)\right\}\\
& =\sqrt{nh^{3}}\left\{\frac{1}{n}\sum_{i=1}^{n}G\left(z, \hat{\gamma}\right)-\frac{1}{n}\sum_{i=1}^{n}G\left(z, \bar{\gamma}\right)-\int G\left(z, \hat{\gamma}\right)dF_{0}\left(z\right)+\int G\left(z, \bar{\gamma}\right)dF_{0}\left(z\right)\right\}\\
& =\sqrt{nh^{3}}\left\{\frac{1}{n^{2}}\sum_{i=1}^{n}\sum_{j=1}^{n}m\left(Z_{i}, Z_{j}\right)-\frac{1}{n}\sum_{i=1}^{n}m_{1}\left(Z_{i}\right)-\frac{1}{n}\sum_{i=1}^{n}m_{2}\left(Z_{i}\right)+E\left[m_{1}\left(z\right)\right]\right\}\\
& = \sqrt{nh^{3}} \times O_{p}\left\{E\left[\abs{m\left(Z_{1}, Z_{1}\right)}\right]/n + \left(E\left[\abs{m\left(Z_{1}, Z_{2}\right)}^{2}\right]\right)^{1/2}/n\right\},
\end{align*}
where the last equality follows from Lemma 8.4 of \citet*{newey1994large}. The last term converges to zero in probability if we can control the convergence rates of $E\left[\abs{m\left(Z_{1}, Z_{1}\right)}\right]$ and $E\left[\abs{m\left(Z_{1}, Z_{2}\right)}^{2}\right]$. Notice that we have $\abs{G\left(z, \gamma\right)}\leq b\left(z\right)\left\lVert \gamma \right\lVert_{2}$ with 

\[
b\left(z\right)=\left\lVert f_{X\mid D=1}\left(x\right)^{-1}\left[1, -f_{Y\mid X, D=1}^{\left(1\right)}\left(\theta_{1}^{*},x\right)\right]\right\lVert_{2}
\] 
where $\left\lVert \cdot \right\lVert_{2}$ denotes the $\ell_{2}$ norm. Then $E\left[\abs{G\left(z, qK_{h}\left(\cdot - x\right)\right)}\right]\leq b\left(z\right)h^{-d}\left\lVert q \right\lVert_{2}$ by the boundedness of $K\left(u\right)$. By that $f_{X\mid D=1}\left(x\right)$ is bounded away from zero and $f_{Y\mid X,D=1}\left(\theta_{1}^{*}, x\right)$ is bounded from above, we have that $E\left[b\left(z\right)^{2}\right]\leq \infty$. Therefore, we have
\begin{align*}
& \sqrt{nh^{3}} \times O_{p}\left\{E\left[\abs{m\left(Z_{1}, Z_{1}\right)}\right]/n + \left(E\left[\abs{m\left(Z_{1}, Z_{2}\right)}^{2}\right]\right)^{1/2}/n\right\}\\
& = \sqrt{nh^{3}} \times O_{p}\left\{E\left[\left\lVert q \right\lVert_{2} b\left(Z_{1}\right)\right]/n + \left(E\left[\left\lVert q \right\lVert_{2}^{2}b\left(Z_{1}\right)^{2}\right]\right)^{1/2}\left(nh^{d}\right)^{-1}\right\}\\
& =\sqrt{nh^{3}}\times O_{p}\left(n^{-1}h^{-d-2}\right)\\
& =o_{p}\left(1\right)
\end{align*}
by the assumptions on $n$ and $h$. The additional $h^{-2}$ in the rates of convergence follows from that $q$ contains  $K_{h}^{\left(1\right)}\left(u\right)=h^{-2}K^{\left(1\right)}\left(u/h\right)$ with bounded $K^{\left(1\right)}\left(u\right)$.

\textit{Step 4.}
In this step, we show that 
\[
\frac{\sqrt{nh^{3}}}{n}\sum_{i=1}^{n}\hat{f}_{Y\mid X,D=1}^{\left(1\right)}\left(\theta_{1}^{*}\mid X_{i}\right)  = \frac{\sqrt{nh^{3}}}{n}\sum_{i=1}^{n}f_{Y\mid X,D=1}^{\left(1\right)}\left(\theta_{1}^{*}\mid X_{i}\right)+\frac{\sqrt{nh^{3}}}{n}\sum_{i=1}^{n} v\left(X_{i}\right) q_{i}+ o_{p}\left(1\right),
\]
where $v\left(X_{i}\right)=\frac{P\left(D=1\right)}{P\left(D=1\mid X_{i}\right)}\left[1,-\frac{\gamma_{10}\left(X_{i}\right)}{\gamma_{20}\left(X_{i}\right)}\right]$ and $q_{i}=\left(\frac{D_{i}K_{h}^{\left(1\right)}\left(\theta_{1}^{*}-Y_{i}\right)}{P\left(D=1\right)}, \frac{D_{i}}{P\left(D=1\right)}\right)'$. To do this, it suffices to show that
\[
\sqrt{nh^{3}}\int G\left(z,\hat{\gamma}-
\gamma_{0}\right)dF_{0}\left(z\right)-\frac{\sqrt{nh^{3}}}{n}\sum_{i=1}^{n} v\left(X_{i}\right) q_{i}=o_{p}\left(1\right)
\]

Notice that

\begin{align*}
\int G\left(z,\gamma\right)dF_{0}\left(z\right)& =\int \gamma_{20}\left(x\right)^{-1}\left[1,-\frac{\gamma_{10}\left(x\right)}{\gamma_{20}\left(x\right)}\right]\gamma\left(x\right)f_{X}\left(x\right)dx\\
& = \int f_{X\mid D=1}\left(x\right)^{-1}\left[1,-\frac{\gamma_{10}\left(x\right)}{\gamma_{20}\left(x\right)}\right]\gamma\left(x\right)f_{X}\left(x\right)dx\\
& = \int \frac{P\left(D=1\right)}{P\left(D=1\mid X=x\right)}\left[1,-\frac{\gamma_{10}\left(x\right)}{\gamma_{20}\left(x\right)}\right]\gamma\left(x\right)dx\\
& = \int v\left(x\right)\gamma\left(x\right)dx,
\end{align*}
where $f_{X}\left(x\right)$ is the density function of $X$ and $v\left(x\right)=\frac{P\left(D=1\right)}{P\left(D=1\mid X=x\right)}\left[1,-\frac{\gamma_{10}\left(x\right)}{\gamma_{20}\left(x\right)}\right]$. Also, we have
\begin{align*}
v\left(x\right)\gamma_{0}\left(x\right)& =\frac{P\left(D=1\right)}{P\left(D=1\mid X=x\right)}\left[1,-\frac{\gamma_{10}\left(x\right)}{\gamma_{20}\left(x\right)}\right]\begin{bmatrix}
           \gamma_{10}\left(x\right) \\
           \gamma_{20}\left(x\right)
         \end{bmatrix}\\
& = \frac{P\left(D=1\right)}{P\left(D=1\mid X=x\right)}\left(\gamma_{10}\left(x\right)-\gamma_{10}\left(x\right)\right)\\
& =0.
\end{align*}
Therefore, we have
\begin{align*}
\int G\left(z,\hat{\gamma}-\gamma_{0}\right)dF_{0}\left(z\right)& = \int v\left(x\right)\hat{\gamma}\left(x\right)dx-\int v\left(x\right)\gamma_{0}\left(x\right)dx=\int v\left(x\right)\hat{\gamma}\left(x\right)dx\\
& = \frac{1}{n}\sum_{i=1}^{n}\int v\left(x\right) q_{i}K_{h}\left(x-X_{i}\right)dx\\
& = \frac{1}{n}\sum_{i=1}^{n} v\left(X_{i}\right) q_{i}+ \left(\frac{1}{n}\sum_{i=1}^{n}\int v\left(x\right) q_{i}K_{h}\left(x-X_{i}\right)dx-\frac{1}{n}\sum_{i=1}^{n} v\left(X_{i}\right) q_{i}\right)\\
& = \frac{1}{n}\sum_{i=1}^{n} v\left(X_{i}\right) q_{i}+ \frac{1}{n}\sum_{i=1}^{n}\left[\int v\left(x\right)K_{h}\left(x-X_{i}\right)dx-v\left(X_{i}\right) \right]q_{i}.
\end{align*}
By Chebyshev's inequality, sufficient conditions for $\sqrt{nh^{3}}$ times the second term in the last line converging to zero in probability are that $$\sqrt{nh^{3}}E\left[\left(\int v\left(x\right)K_{h}\left(x-X_{i}\right)dx-v\left(X_{i}\right) \right)q_{i}\right]\to 0$$and
$$E\left[\left\lVert q_{i}\right\lVert^{2} \left\lVert\int v\left(x\right)K_{h}\left(x-X_{i}\right)dx-v\left(X_{i}\right)\right\lVert^{2} \right]\to 0.$$
The expectation in the first condition is the difference of  $E\left[\left(\int v\left(x\right)K_{h}\left(x-X_{i}\right)dx \right)q_{i}\right]$ and $E\left[v\left(X_{i}\right)q_{i}\right]$. We begin with the second term $E\left[v\left(X_{i}\right)q_{i}\right]$. Notice that
\begin{align*}
E\left[v\left(X_{i}\right)q_{i}\right]& =E\left[v\left(X_{i}\right)E\left[q_{i}\mid X_{i}\right]\right]\\
& = E\left[v\left(X_{i}\right)E\left[\begin{pmatrix}
\frac{D_{i}K_{h}^{\left(1\right)}\left(\theta_{1}^{*}-Y_{i}\right)}{P\left(D=1\right)}\\
\frac{D_{i}}{P\left(D=1\right)}
\end{pmatrix}\mid X_{i}\right]\right]\\
& = E\left[v\left(X_{i}\right)\frac{P\left(D=1\mid X_{i}\right)}{P\left(D=1\right)}E\left[\begin{pmatrix}
K_{h}^{\left(1\right)}\left(\theta_{1}^{*}-Y_{i}\right)\\
1
\end{pmatrix}\mid X_{i}, D=1\right]\right]
\end{align*}
by the law of iterated expectations. 
The inner conditional expectation in the last line satisfies 

\begin{align*}
E\left[K_{h}^{\left(1\right)}\left(\theta_{1}^{*}-Y_{i}\right)\mid X_{i}, D=1\right]& = \frac{1}{h^{2}}E\left[K^{\left(1\right)}\left(\frac{\theta_{1}^{*}-Y_{i}}{h}\right)\mid X_{i}, D=1\right]\\
& = \frac{1}{h^{2}}\int K^{\left(1\right)}\left(\frac{\theta_{1}^{*}-y}{h}\right) f_{Y\mid X, D=1}\left(y\mid X_{i}\right)dy\\
& = \frac{1}{h} \int  K\left(\frac{\theta_{1}^{*}-y}{h}\right) f_{Y\mid X, D=1}^{\left(1\right)}\left(y\mid X_{i}\right)dy\\
& = \int K\left(u\right)f_{Y\mid X, D=1}^{\left(1\right)}\left(\theta_{1}^{*}+hu\mid X_{i}\right)du\\
& = \int K\left(u\right)f_{Y\mid X, D=1}^{\left(1\right)}\left(\theta_{1}^{*}\mid X_{i}\right)du\\
& +\int huK\left(u\right)f_{Y\mid X, D=1}^{\left(1\right)}\left(\theta_{1}^{*}\mid X_{i}\right)du\\
& + \int \frac{h^{2}u^{2}}{2} K\left(u\right)f_{Y\mid X, D=1}^{\left(3\right)}\left(\tilde{\theta}_{1}\mid X_{i}\right)du\\
& = f_{Y\mid X, D=1}^{\left(1\right)}\left(\theta_{1}^{*}\mid X_{i}\right)+ \frac{h^{2}}{2}\kappa_{2}f_{Y\mid X, D=1}^{\left(3\right)}\left(\tilde{\theta}_{1}\mid X_{i}\right)
\end{align*}
with $\tilde{\theta}_{1}\in \left(\theta_{1}^{*}, \theta_{1}^{*}+hu\right)$ and $\kappa_{2}=\int u^{2}K\left(u\right)du$. The third equality follows from integration by parts and the forth from change of variables. Hence, 

\begin{align*}
E\left[v\left(X_{i}\right)q_{i}\right]& = E\left[v\left(X_{i}\right)\frac{P\left(D=1\mid X_{i}\right)}{P\left(D=1\right)}\begin{pmatrix}
f_{Y\mid X, D=1}^{\left(1\right)}\left(\theta_{1}^{*}\mid X_{i}\right)\\
1
\end{pmatrix}\right]\\
& + \frac{h^{2}}{2}\kappa_{2}E\left[v\left(X_{i}\right)\frac{P\left(D=1\mid X_{i}\right)}{P\left(D=1\right)}\begin{pmatrix}
f_{Y\mid X, D=1}^{\left(3\right)}\left(\tilde{\theta}_{1}\mid X_{i}\right)\\
0
\end{pmatrix}\right]\\ 
& = E\left[v\left(X_{i}\right)\frac{P\left(D=1\mid X_{i}\right)}{P\left(D=1\right)}\begin{pmatrix}
f_{Y\mid X, D=1}^{\left(1\right)}\left(\theta_{1}^{*}\mid X_{i}\right)\\
1
\end{pmatrix}\right]+ O\left(h^{2}\right)\\
& = \int v\left(x\right)\frac{P\left(D=1\mid X_{i}=x\right)}{P\left(D=1\right)}\begin{pmatrix}
f_{Y\mid X, D=1}^{\left(1\right)}\left(\theta_{1}^{*}\mid X_{i}=x\right)\\
1
\end{pmatrix}f_{X}\left(x\right)dx+O\left(h^{2}\right)\\
& = \int v\left(x\right)\begin{pmatrix}
f_{Y\mid X, D=1}^{\left(1\right)}\left(\theta_{1}^{*}\mid X_{i}=x\right)\\
1
\end{pmatrix}f_{X\mid D=1}\left(x\right)dx+O\left(h^{2}\right)\\
& = \int v\left(x\right)\begin{pmatrix}
f_{Y,X\mid D=1}^{\left(1\right)}\left(\theta_{1}^{*}\mid X_{i}=x\right)\\
f_{X\mid D=1}\left(x\right)
\end{pmatrix}dx+O\left(h^{2}\right)\\
& = \int v\left(x\right)\gamma_{0}\left(x\right)dx+O\left(h^{2}\right).
\end{align*}
Using the same arguments,  we can also show that 
\begin{align*}
E\left[\left(\int v\left(x\right)K_{h}\left(x-X_{i}\right)dx \right)q_{i}\right]& =E\left[\left(\int v\left(X_{i}+hu\right)K\left(u\right)du \right)q_{i}\right]\\
& =\int \left(\int v\left(x+hu\right)K\left(u\right)du \right)\gamma_{0}\left(x\right)dx+O\left(h^{2}\right)
\end{align*}
Then the first condition equals
\begin{align*}
& \sqrt{nh^{3}}\left\lVert E\left[\left(\int v\left(x\right)K_{h}\left(x-X_{i}\right)dx-v\left(X_{i}\right) \right)q_{i}\right]\right\lVert\\
& = \sqrt{nh^{3}}\left\lVert \int \left(\int v\left(x+hu\right)K\left(u\right)du \right)\gamma_{0}\left(x\right)dx - \int v\left(x\right)\gamma_{0}\left(x\right)dx+O\left(h^{2}\right)\right\lVert.
\end{align*}
Following the argument in Theorem 8.11 of \citet*{newey1994large}, the last line satisfies
\begin{align*}
& \sqrt{nh^{3}}\left\lVert \int \int v\left(x\right)K\left(u\right) \gamma_{0}\left(x-hu\right)dudx- \int v\left(x\right)\gamma_{0}\left(x\right)dx+O\left(h^{2}\right)\right\lVert\\
& = \sqrt{nh^{3}}\left\lVert \int v\left(x\right)\left\{\int \left[\gamma_{0}\left(x-hu\right)-\gamma_{0}\left(x\right)\right]du\right\}dx +O\left(h^{2}\right)\right\lVert\\
& \leq \sqrt{nh^{3}} \int \left\lVert v\left(x\right) \right\lVert \left\lVert \int \left[\gamma_{0}\left(x-hu\right)-\gamma_{0}\left(x\right)\right]du\right\lVert dx+ O\left(\sqrt{nh^{3}}h^{2}\right)\\
& \leq \sqrt{nh^{3}}Ch^{2}\int \left\lVert v\left(x\right) \right\lVert dx+ O\left(\sqrt{nh^{3}}h^{2}\right)\\
& = O\left(\sqrt{nh^{3}}h^{2}\right).
\end{align*}
Therefore the first condition holds if $\sqrt{nh^{3}}h^{2}\to 0$. 

Recall that the second condition we would like to show is 

\[
E\left[\left\lVert q_{i}\right\lVert^{2} \left\lVert\int v\left(x\right)K_{h}\left(x-X_{i}\right)dx-v\left(X_{i}\right)\right\lVert^{2} \right]\to 0.
\]
By Cauchy Schwartz inequality, it suffices to show that 
\[E\left[ \left\lVert\int v\left(x\right)K_{h}\left(x-X_{i}\right)dx-v\left(X_{i}\right)\right\lVert^{4} \right]\to 0.
\]
By the continuity of $v\left(x\right)$, $v\left(x+hu\right)\to v\left(x\right)$ for all $x$ and $u$ as $h\to 0$. By the dominated convergence theorem, $\int v\left(x\right)K_{h}\left(x-x_{i}\right)dx=\int v\left(x+hu\right)K\left(u\right)du\to \int v\left(x\right)K\left(u\right)du=v\left(x\right)$ for all $x$. Therefore we have 
\begin{align*}
& E\left[ \left\lVert\int v\left(x\right)K_{h}\left(x-X_{i}\right)dx-v\left(X_{i}\right)\right\lVert^{4} \right]= E\left[ \left\lVert\int v\left(X_{i}+hu\right)K\left(u\right)du-v\left(X_{i}\right)\right\lVert^{4} \right]\to 0.
\end{align*}

\textit{Step 5.}
By Step 4 and the definition of $v\left(X_{i}\right)$ and $q_{i}$, we have 
\begin{align*}
\frac{\sqrt{nh^{3}}}{n}\sum_{i=1}^{n}\hat{f}_{Y\mid X, D=1}^{\left(1\right)}\left(\theta_{1}^{*}\mid X_{i}\right)& =\frac{\sqrt{nh^{3}}}{n}\sum_{i=1}^{n}f_{Y\mid X, D=1}^{\left(1\right)}\left(\theta_{1}^{*}\mid X_{i}\right)+\frac{\sqrt{nh^{3}}}{n}\sum_{i=1}^{n}v\left(X_{i}\right)q_{i}+o_{p}\left(1\right)\\
& = \frac{\sqrt{nh^{3}}}{n}\sum_{i=1}^{n}f_{Y\mid X, D=1}^{\left(1\right)}\left(\theta_{1}^{*}\mid X_{i}\right)+o_{p}\left(1\right)\\
& + \frac{\sqrt{nh^{3}}}{n}\sum_{i=1}^{n}\frac{D_{i}}{P\left(D=1\mid X_{i}\right)}\left[K_{h}^{\left(1\right)}\left(\theta_{1}^{*}-Y_{i}\right)-f_{Y\mid X, D=1}^{\left(1\right)}\left(\theta_{1}^{*}\mid X_{i}\right)\right]\\
& = \frac{\sqrt{nh^{3}}}{n}\sum_{i=1}^{n}f_{Y\mid X, D=1}^{\left(1\right)}\left(\theta_{1}^{*}\mid X_{i}\right)\left[1-\frac{D_{i}}{P\left(D=1\mid X_{i}\right)}\right]\\
& + \frac{\sqrt{nh^{3}}}{n}\sum_{i=1}^{n}\frac{D_{i}}{P\left(D=1\mid X_{i}\right)}K_{h}^{\left(1\right)}\left(\theta_{1}^{*}-Y_{i}\right)+o_{p}\left(1\right).
\end{align*}
Since  we have $
E\left[f_{Y\mid X, D=1}^{\left(1\right)}\left(\theta_{1}^{*}\mid X_{i}\right)\left[1-\frac{D_{i}}{P\left(D=1\mid X_{i}\right)}\right]\right]= 0
$ by the law of iterated expectations, the central limit theorem holds for the first term of r.h.s. Hence,

\begin{align*}
\frac{\sqrt{nh^{3}}}{n}\sum_{i=1}^{n}\hat{f}_{Y\mid X, D=1}^{\left(1\right)}\left(\theta_{1}^{*}\mid X_{i}\right)& = O_{p}\left(\sqrt{nh^{3}}n^{-1/2}\right)+\frac{\sqrt{nh^{3}}}{nh^{2}}\sum_{i=1}^{n}\frac{D_{i}}{P\left(D=1\mid X_{i}\right)}K^{\left(1\right)}\left(\frac{\theta_{1}^{*}-Y_{i}}{h}\right)\\
& +o_{p}\left(1\right)\\
& = O_{p}\left(\sqrt{h^{3}}\right)+\frac{1}{\sqrt{nh}}\sum_{i=1}^{n}\frac{D_{i}}{P\left(D=1\mid X_{i}\right)}K^{\left(1\right)}\left(\frac{\theta_{1}^{*}-Y_{i}}{h}\right)+o_{p}\left(1\right)\\
& =\frac{1}{\sqrt{nh}}\sum_{i=1}^{n}\frac{D_{i}}{P\left(D=1\mid X_{i}\right)}K^{\left(1\right)}\left(\frac{\theta_{1}^{*}-Y_{i}}{h}\right)+o_{p}\left(1\right).
\end{align*}
In this step, we show that  
\[
\frac{1}{\sqrt{nh}}\sum_{i=1}^{n}\frac{D_{i}}{P\left(D=1\mid X_{i}\right)}K^{\left(1\right)}\left(\frac{\theta_{1}^{*}-Y_{i}}{h}\right)\overset{d}\to N\left(0,V\right),
\]
where $V=\kappa_{0}^{\left(1\right)}E\left[\frac{f_{Y\mid X, D=1}\left(\theta_{1}^{*}\mid X\right)}{P\left(D=1\mid X\right)}\right]$ and $\kappa_{0}^{\left(1\right)}=\int K^{\left(1\right)}\left(u\right)^{2}du$.

For convenience, we define $\hat{g}\left(\theta_{1}^{*}\right)\equiv \left(nh^{2}\right)^{-1}\sum_{i=1}^{n}\frac{D_{i}}{P\left(D=1\mid X_{i}\right)}K^{\left(1\right)}\left(\frac{\theta_{1}^{*}-Y_{i}}{h}\right)$. Then it is equivalent to show that
\[
\sqrt{nh^{3}}\left(\hat{g}\left(\theta_{1}^{*}\right)-0\right)\overset{d}\to N\left(0,V\right).
\]
To use central limit theorem, we have to calculate $E\left[\hat{g}\left(\theta_{1}^{*}\right)\right]$ and $Var\left(\hat{g}\left(\theta_{1}^{*}\right)\right)$. 
\begin{align*}
E\left[\hat{g}\left(\theta_{1}^{*}\right)\right]& =\frac{1}{h^{2}}E\left[\frac{D_{i}}{P\left(D=1\mid X_{i}\right)}K^{\left(1\right)}\left(\frac{\theta_{1}^{*}-Y_{i}}{h}\right)\right]\\
& =\frac{1}{h^{2}}E\left[\frac{1}{P\left(D=1\mid X_{i}\right)}E\left[D_{i}K^{\left(1\right)}\left(\frac{\theta_{1}^{*}-Y_{i}}{h}\right)\mid X_{i}\right]\right]\\
& = \frac{1}{h^{2}} E\left[E\left[K^{\left(1\right)}\left(\frac{\theta_{1}^{*}-Y_{i}}{h}\right)\mid X_{i}, D=1\right]\right].
\end{align*}
Since $h^{-2}E\left[K^{\left(1\right)}\left(\frac{\theta_{1}^{*}-Y_{i}}{h}\right)\mid X_{i}, D=1\right]=f_{Y\mid X, D=1}^{\left(1\right)}\left(\theta_{1}^{*}\mid X_{i}\right)+\frac{h^{2}}{2}\kappa_{2} f_{Y\mid X, D=1}^{\left(3\right)}\left(\tilde{\theta}_{1}\mid X_{i}\right)$ from the calculation in Step 4, then
\begin{align*}
E\left[\hat{g}\left(\theta_{1}^{*}\right)\right]=E\left[f_{Y\mid X, D=1}^{\left(1\right)}\left(\theta_{1}^{*}\mid X_{i}\right)\right]+O\left(h^{2}\right)=0+O\left(h^{2}\right).
\end{align*}
For the variance, 
\begin{align*}
Var\left(\hat{g}\left(\theta_{1}^{*}\right)\right)& =\frac{1}{nh^{4}}Var\left(\frac{D_{i}}{P\left(D=1\mid X_{i}\right)}K^{\left(1\right)}\left(\frac{\theta_{1}^{*}-Y_{i}}{h}\right)\right)\\
& = \frac{1}{nh^{4}}E\left[\left(\frac{D_{i}}{P\left(D=1\mid X_{i}\right)}K^{\left(1\right)}\left(\frac{\theta_{1}^{*}-Y_{i}}{h}\right)\right)^2\right]\\
& - \frac{1}{nh^{4}}\left(E\left[\frac{D_{i}}{P\left(D=1\mid X_{i}\right)}K^{\left(1\right)}\left(\frac{\theta_{1}^{*}-Y_{i}}{h}\right)\right]\right)^{2}\\
& = \frac{1}{nh^{4}}E\left[\left(\frac{D_{i}}{P\left(D=1\mid X_{i}\right)}K^{\left(1\right)}\left(\frac{\theta_{1}^{*}-Y_{i}}{h}\right)\right)^2\right]+ \frac{1}{nh^{4}}O\left(h^{4}\right)\\
& = \frac{1}{nh^{4}}E\left[\frac{1}{P\left(D=1\mid X_{i}\right)^2}E\left[D_{i}^2K^{\left(1\right)}\left(\frac{\theta_{1}^{*}-Y_{i}}{h}\right)^2\mid X\right]\right]+\frac{1}{nh^{4}}O\left(h^{4}\right)\\
& = \frac{1}{nh^{4}}E\left[\frac{1}{P\left(D=1\mid X_{i}\right)}E\left[K^{\left(1\right)}\left(\frac{\theta_{1}^{*}-Y_{i}}{h}\right)^2\mid X, D=1\right]\right]+\frac{1}{nh^{4}}O\left(h^{4}\right).
\end{align*}
The inner expectation in the last line equals

\begin{align*}
E\left[K^{\left(1\right)}\left(\frac{\theta_{1}^{*}-Y_{i}}{h}\right)^2\mid X, D=1\right]& = \int K^{\left(1\right)}\left(\frac{\theta_{1}^{*}-y}{h}\right)^2 f_{Y\mid X, D=1}\left(y\mid X\right)dy\\
& = h\int K^{\left(1\right)}\left(u\right)^2 f_{Y\mid X, D=1}\left(\theta_{1}^{*}+hu\mid X\right)du\\
& = hf_{Y\mid X, D=1}\left(\theta_{1}^{*}\mid X\right)\int K^{\left(1\right)}\left(u\right)^2du\\
& + h^{2}f_{Y\mid X, D=1}^{\left(1\right)}\left(\tilde{\theta}_{1}\mid X\right)\int uK^{\left(1\right)}\left(u\right)^2du,
\end{align*}
where $\tilde{\theta}_{1}\in \left(\theta_{1}^{*},\theta_{1}^{*}+hu\right)$. Define $\kappa_{0}^{\left(1\right)}=\int K^{\left(1\right)}\left(u\right)^2du$ and $\kappa_{1}^{\left(1\right)}=\int u K^{\left(1\right)}\left(u\right)^2du$, the variance equals

\begin{align*}
Var\left(\hat{g}\left(\theta_{1}^{*}\right)\right)& = \frac{1}{nh^{4}}E\left[\frac{h}{P\left(D=1\mid X_{i}\right)}\kappa_{0}^{\left(1\right)}f_{Y\mid X, D=1}\left(\theta_{1}^{*}\mid X\right)\right]\\
& +\frac{1}{nh^{4}}E\left[\frac{h^{2}}{P\left(D=1\mid X_{i}\right)}\kappa_{1}^{\left(1\right)}f_{Y\mid X, D=1}^{\left(1\right)}\left(\theta_{1}^{*}\mid X\right)\right]+\frac{1}{nh^{4}}O\left(h^{4}\right)\\
& = \frac{1}{nh^{3}}\left(\kappa_{0}^{\left(1\right)}E\left[\frac{1}{P\left(D=1\mid X_{i}\right)}f_{Y\mid X, D=1}\left(\theta_{1}^{*}\mid X\right)\right]+O\left(h\right)+O\left(h^{3}\right)\right)\\
& =\frac{1}{nh^{3}}\left(V+O\left(h\right)+O\left(h^{3}\right)\right). 
\end{align*}
Then we are ready to apply the central limit theorem. 

Let 
\[
Z_{n,i}\equiv \left(nh\right)^{-1/2}\left(\frac{D_{i}}{P\left(D=1\mid X_{i}\right)}K^{\left(1\right)}\left(\frac{\theta_{1}^{*}-Y_{i}}{h}\right)-E\left[\frac{D_{i}}{P\left(D=1\mid X_{i}\right)}K^{\left(1\right)}\left(\frac{\theta_{1}^{*}-Y_{i}}{h}\right)\right]\right),
\] 
then $E\left[Z_{n,i}\right]=0$ and $Var\left(Z_{n,i}\right)=h^{3}Var\left(\hat{g}\left(\theta_{1}^{*}\right)\right)=n^{-1}V+o\left(n^{-1}\right)$. Then 
\begin{align*}
\sqrt{nh^{3}}\left(\hat{g}\left(\theta_{1}^{*}\right)-0\right)& =\sqrt{nh^{3}}\left(\hat{g}\left(\theta_{1}^{*}\right)-E\left[\hat{g}\left(\theta_{1}^{*}\right)\right]\right)+\sqrt{nh^{3}}\left(E\left[\hat{g}\left(\theta_{1}^{*}\right)\right]-0\right)\\
& =\sqrt{nh^{3}}\left(\hat{g}\left(\theta_{1}^{*}\right)-E\left[\hat{g}\left(\theta_{1}^{*}\right)\right]\right)+ \sqrt{nh^{3}}O\left(h^{2}\right)\\
& = \sum_{i=1}^{n}Z_{n,i}+\sqrt{nh^{3}}O\left(h^{2}\right)\\
& \overset{d}\to N\left(0,V\right)
\end{align*}
by Liapunov CLT and $\sqrt{nh^{3}}h^{2}\to 0$. 

\textit{Step 6.} In this step, we show that 
\[
\sqrt{nh^{3}}\begin{bmatrix}
\hat{\theta}_{1}-\theta_{1}^{*}\\
\hat{\theta}_{0}-\theta_{0}^{*}
\end{bmatrix}\overset{d}{\to}N\left(\begin{bmatrix}
0\\
0
\end{bmatrix}, \begin{bmatrix}
M_{1}V_{1}M_{1} & 0\\
0 & M_{0}V_{0}M_{0}
\end{bmatrix}\right)
\]
and thus, by the delta method, we have
\[
\sqrt{nh^{3}}\left(\hat{\Delta}-\Delta^{*}\right)\overset{d}{\to}N\left(0,M_{1}V_{1}M_{1}+M_{0}V_{0}M_{0}\right).
\]

To show the joint distribution we adopt vector notations. The first-order conditions of $\hat{\theta}_{1}$ and $\hat{\theta}_{}$ give

\begin{align*}
\begin{bmatrix}
0\\
0
\end{bmatrix}& =\begin{bmatrix}
\hat{f}_{Y_{1}}\left(\hat{\theta}_{1}\right)\\
\hat{f}_{Y_{0}}\left(\hat{\theta}_{0}\right)
\end{bmatrix}=\frac{1}{n}\sum_{i=1}^{n}\begin{bmatrix}
\hat{f}_{Y\mid X, D=1}^{\left(1\right)}\left(\hat{\theta}_{1}\mid X_{i}\right)\\
\hat{f}_{Y\mid X, D=0}^{\left(1\right)}\left(\hat{\theta}_{0}\mid X_{i}\right)
\end{bmatrix} = \frac{1}{n}\sum_{i=1}^{n}\begin{bmatrix}
\hat{f}_{Y\mid X, D=1}^{\left(1\right)}\left(\theta_{1}^{*}\mid X_{i}\right)\\
\hat{f}_{Y\mid X, D=0}^{\left(1\right)}\left(\theta_{0}^{*}\mid X_{i}\right)
\end{bmatrix}+J_{n}\begin{bmatrix}
\hat{\theta}_{1}-\theta_{1}^{*}\\
\hat{\theta}_{0}-\theta_{0}^{*}
\end{bmatrix}
\end{align*}
where

\begin{align*}
J_{n}& =\frac{1}{n}\sum_{i=1}^{n}\begin{bmatrix}
\frac{\partial \hat{f}_{Y\mid X, D=1}^{\left(1\right)}\left(\tilde{\theta}_{1}\mid X_{i}\right)}{\partial \theta_{1}} & \frac{\partial \hat{f}_{Y\mid X, D=1}^{\left(1\right)}\left(\tilde{\theta}_{1}\mid X_{i}\right)}{\partial \theta_{0}}\\
\frac{\partial \hat{f}_{Y\mid X, D=0}^{\left(1\right)}\left(\tilde{\theta}_{0}\mid X_{i}\right)}{\partial \theta_{1}} & \frac{\partial \hat{f}_{Y\mid X, D=0}^{\left(1\right)}\left(\tilde{\theta}_{0}\mid X_{i}\right)}{\partial \theta_{0}}
\end{bmatrix}\\
& =\frac{1}{n}\sum_{i=1}^{n}\begin{bmatrix}
\hat{f}_{Y\mid X, D=1}^{\left(2\right)}\left(\tilde{\theta}_{1}\mid X_{i}\right) & 0\\
0 & \hat{f}_{Y\mid X, D=0}^{\left(2\right)}\left(\tilde{\theta}_{0}\mid X_{i}\right)
\end{bmatrix}.
\end{align*}
Hence we have 
\begin{align*}
\sqrt{nh^{3}}\begin{bmatrix}
\hat{\theta}_{1}-\theta_{1}^{*}\\
\hat{\theta}_{0}-\theta_{0}^{*}
\end{bmatrix}& =\left(J_{n}^{-1}\right)\frac{\sqrt{nh^{3}}}{n}\sum_{i=1}^{n}\begin{bmatrix}
\hat{f}_{Y\mid X, D=1}^{\left(1\right)}\left(\hat{\theta}_{1}\mid X_{i}\right)\\
\hat{f}_{Y\mid X, D=0}^{\left(1\right)}\left(\hat{\theta}_{0}\mid X_{i}\right)
\end{bmatrix}\\
& = \left(J_{n}^{-1}\right)\frac{1}{\sqrt{nh}}\sum_{i=1}^{n}\begin{bmatrix}
\frac{D_{i}}{P\left(D=1\mid X_{i}\right)}K^{\left(1\right)}\left(\frac{\theta_{1}^{*}-Y_{i}}{h}\right)\\
\frac{1-D_{i}}{P\left(D=0\mid X_{i}\right)}K^{\left(1\right)}\left(\frac{\theta_{0}^{*}-Y_{i}}{h}\right)
\end{bmatrix}+\begin{bmatrix}
o_{p}\left(1\right)\\
o_{p}\left(1\right)
\end{bmatrix}
\end{align*}
where the last equality follows from the Step 5 in the proof of Theorem 1. Since $$J_{n}\overset{p}{\to}\begin{bmatrix}
M_{1} & 0\\
0 & M_{0}
\end{bmatrix}$$ and 

\[
\frac{1}{\sqrt{nh}}\sum_{i=1}^{n}\begin{bmatrix}
\frac{D_{i}}{P\left(D=1\mid X_{i}\right)}K^{\left(1\right)}\left(\frac{\theta_{1}^{*}-Y_{i}}{h}\right)\\
\frac{1-D_{i}}{P\left(D=0\mid X_{i}\right)}K^{\left(1\right)}\left(\frac{\theta_{0}^{*}-Y_{i}}{h}\right)
\end{bmatrix}\overset{d}{\to}N\left(\begin{bmatrix}
0\\
0
\end{bmatrix}, \begin{bmatrix}
V_{1}& 0\\
0 & V_{0}
\end{bmatrix}\right),
\]
then by Slutsky's theorem we have
\[
\sqrt{nh^{3}}\begin{bmatrix}
\hat{\theta}_{1}-\theta_{1}^{*}\\
\hat{\theta}_{0}-\theta_{0}^{*}
\end{bmatrix}\overset{d}{\to}N\left(\begin{bmatrix}
0\\
0
\end{bmatrix}, \begin{bmatrix}
M_{1}V_{1}M_{1} & 0\\
0 & M_{0}V_{0}M_{0}
\end{bmatrix}\right).
\]
\vskip 1cm

\textit{Proof of Theorem 3.} It is enough to show the results of $\hat{M}_{1}$ and $\hat{V}_{1}$. We first show that $\hat{M}_{1}\overset{p}\to M_{1}$. By adding and subtracting additional terms, we have

\[
\hat{M}_{1}=\frac{1}{n}\sum_{i=1}^{n}\hat{f}_{Y\mid X, D=1}^{\left(2\right)}\left(\hat{\theta}_{1}\mid X_{i}\right)=\frac{1}{n}\sum_{i=1}^{n}f_{Y\mid X, D=1}^{\left(2\right)}\left(\theta_{1}^{*}\mid X_{i}\right)+A_{1}+A_{2}
\]
where 
\[
A_{1}=\frac{1}{n}\sum_{i=1}^{n}\hat{f}_{Y\mid X, D=1}^{\left(2\right)}\left(\hat{\theta}_{1}\mid X_{i}\right)-f_{Y\mid X, D=1}^{\left(2\right)}\left(\hat{\theta}_{1}\mid X_{i}\right)
\]
and
\[
A_{2}=\frac{1}{n}\sum_{i=1}^{n}f_{Y\mid X, D=1}^{\left(2\right)}\left(\hat{\theta}_{1}\mid X_{i}\right)-f_{Y\mid X, D=1}^{\left(2\right)}\left(\theta_{1}^{*}\mid X_{i}\right).
\]
If we can show that $A_{1}=o_{p}\left(1\right)$ and $A_{2}=o_{p}\left(1\right)$, then $\hat{M}\overset{p}\to M$ by law of large numbers. Note that 
\begin{align*}
\abs{A_{1}}& \leq \frac{1}{n}\sum_{i=1}^{n}\abs{\hat{f}_{Y\mid X, D=1}^{\left(2\right)}\left(\hat{\theta}_{1}\mid X_{i}\right)-f_{Y\mid X, D=1}^{\left(2\right)}\left(\hat{\theta}_{1}\mid X_{i}\right)}\\
& \leq \sup_{y,x}\abs{\hat{f}_{Y\mid X, D=1}^{\left(2\right)}\left(y\mid x\right)-f_{Y\mid X, D=1}^{\left(2\right)}\left(y\mid x\right)}\\
& = O_{p}\left(\sqrt{\frac{\ln{n}}{nh^{d+5}}}+h^{2}\right)\\
& = o_{p}\left(1\right),
\end{align*}
where the first equality follows from the uniform rates of convergence of kernel estimators \citep*{hansen2008uniform}. For $A_{2}$, we use the argument in Lemma 4.3 of \citet*{newey1994large}. By consistency of $\hat{\theta}_{1}$ there is $\delta_{n}\to 0$ such that $\norm{\hat{\theta}_{1}-\theta_{1}^{*}}\leq \delta_{n}$ with probability approaching to one. Define $\Delta_{n}\left(Z_{i}\right)=\sup_{\norm{y-\theta_{1}^{*}}\leq \delta_{n}}\norm{f_{Y\mid X, D=1}^{\left(2\right)}\left(y\mid X_{i}\right)-f_{Y\mid X, D=1}^{\left(2\right)}\left(\theta_{1}^{*}\mid X_{i}\right)}$. By the continuity of $f_{Y\mid X, D=1}^{\left(2\right)}\left(y\mid X_{i}\right)$ at $\theta_{1}^{*}$, $\Delta_{n}\left(Z_{i}\right)\overset{p}{\to} 0$. By the dominated convergence theorem, we have $E\left[\Delta_{n}\left(Z_{i}\right)\right]\to 0$. By Markov inequality, $P\left(n^{-1}\sum_{i=1}^{n}\Delta_{n}\left(Z_{i}\right)> \epsilon\right)\leq E\left[\Delta_{n}\left(Z_{i}\right)\right]/\epsilon\to 0$. Therefore, we have 
\[\abs{A_{2}}\leq \frac{1}{n}\sum_{i=1}^{n}\Delta_{n}\left(Z_{i}\right)=o_{p}\left(1\right).
\] 

Next we show that $\hat{V}_{1}\overset{p}{\to}V_{1}$. We can rewrite 
\[
\hat{V}_{1}/\kappa_{0}^{\left(1\right)}=\frac{1}{n}\sum_{i=1}^{n}\frac{\hat{f}_{Y\mid X,D=1}\left(\hat{\theta}_{1}\mid X_{i}\right)}{\hat{\pi}\left(X_{i}\right)}=\frac{1}{n}\sum_{i=1}^{n}\frac{f_{Y\mid X,D=1}\left(\theta_{1}^{*}\mid X_{i}\right)}{\pi\left(X_{i}\right)}+B_{1}+B_{2}
\]
with 

\[
B_{1}=\frac{1}{n}\sum_{i=1}^{n}\frac{\hat{f}_{Y\mid X,D=1}\left(\hat{\theta}_{1}\mid X_{i}\right)}{\hat{\pi}\left(X_{i}\right)}-\frac{f_{Y\mid X,D=1}\left(\hat{\theta}_{1}\mid X_{i}\right)}{\pi\left(X_{i}\right)}
\]
and
\[
B_{2}=\frac{1}{n}\sum_{i=1}^{n}\frac{f_{Y\mid X,D=1}\left(\hat{\theta}_{1}\mid X_{i}\right)}{\pi\left(X_{i}\right)}-\frac{f_{Y\mid X,D=1}\left(\theta_{1}^{*}\mid X_{i}\right)}{\pi\left(X_{i}\right)}.
\]
It remains to show that $B_{1}=o_{p}\left(1\right)$ and $B_{2}=o_{p}\left(1\right)$. The result of $B_{2}$ follows from the same arguments as in the proof of $A_{2}$ if $f_{Y\mid X,D=1}\left(y\mid X_{i}\right)$ is continuous at $\theta_{1}^{*}$. Thus, we only focus on $B_{1}$. For conenience, define $f\left(y\mid x\right)=f_{Y\mid X,D=1}\left(y\mid x\right)$. For $\pi$ bounded away from zero, we have 

\begin{align*}
\frac{\hat{f}\left(y\mid x\right)}{\hat{\pi}\left(x\right)}-\frac{f\left(y\mid x\right)}{\pi\left(x\right)}& = \frac{\pi\left(x\right)\hat{f}\left(y\mid x\right)-\hat{\pi}\left(x\right)f\left(y\mid x\right)}{\hat{\pi}\left(x\right)\pi\left(x\right)}\\ 
& = \frac{\pi\left(x\right)\hat{f}\left(y\mid x\right)-\pi\left(x\right)f\left(y\mid x\right)+\pi\left(x\right)f\left(y\mid x\right)-\hat{\pi}\left(x\right)f\left(y\mid x\right)}{\hat{\pi}\left(x\right)\pi\left(x\right)}\\
& = \frac{\hat{f}\left(y\mid x\right)-f\left(y\mid x\right)}{\hat{\pi}\left(x\right)}+ \frac{f\left(y\mid x\right)}{\hat{\pi}\left(x\right)\pi\left(x\right)}\left(\hat{\pi}\left(x\right)-\pi\left(x\right)\right)\\
& \leq C\left(\left(\hat{f}\left(y\mid x\right)-f\left(y\mid x\right)\right)+\left(\hat{\pi}\left(x\right)-\pi\left(x\right)\right)\right)
\end{align*}
for some $C>0$. By the uniform rates of convergence of kernel estimators \citep*{hansen2008uniform}, we have
\begin{align*}
\abs{B_{1}}& \leq \frac{1}{n}\sum_{i=1}^{n}\abs{\frac{\hat{f}_{Y\mid X,D=1}\left(\hat{\theta}_{1}\mid X_{i}\right)}{\hat{\pi}\left(X_{i}\right)}-\frac{f_{Y\mid X,D=1}\left(\hat{\theta}_{1}\mid X_{i}\right)}{\pi\left(X_{i}\right)}}\\
& \leq C\sup_{y,x}\left(\abs{\hat{f}\left(y\mid x\right)-f\left(y\mid x\right)}+\abs{\hat{\pi}\left(x\right)-\pi\left(x\right)}\right)\\
& = O_{p}\left(\sqrt{\frac{\ln{n}}{nh^{d+1}}}+h^{2}\right)+\sup_{x}\abs{\hat{\pi}\left(x\right)-\pi\left(x\right)}\\
& = o_{p}\left(1\right)
\end{align*}
by the rates of $n$ and $h$  and the uniform convergence of $\hat{\pi}\left(x\right)$. 

\textit{Proof of Theorem 4:} Suppose that 
\[
\hat{f}_{Y_{1}}(y)=\frac{1}{K}\sum_{k=1}^{K}\mathbb{E}_{n,k}[m_{1}(Z,y,\hat{\eta}_{1k})]
\] 
is differentiable with respect to $y$. Define 
\[
\hat{f}_{Y_{1}}^{(1)}(y)\equiv \frac{1}{K}\sum_{k=1}^{K}\mathbb{E}_{n,k}[m_{1}^{(1)}(Z,y,\hat{\eta}_{1k})]
\]
where $m_{1}^{(1)}(Z,y,\hat{\eta}_{1k})\equiv \partial m_{1}(Z,y,\hat{\eta}_{1k})/ \partial y$.

By the definition of $\hat{\theta}_{1}$, we have

\begin{align*}
0 & = \hat{f}_{Y_{1}}^{(1)}(\hat{\theta}_{1})= \frac{1}{K}\sum_{k=1}^{K}\mathbb{E}_{n,k}[m_{1}^{(1)}(Z,\hat{\theta}_{1},\hat{\eta}_{1k})]\\
& = \frac{1}{K}\sum_{k=1}^{K}\mathbb{E}_{n,k}[m_{1}^{(1)}(Z,\theta_{1}^{*},\hat{\eta}_{1k})]+\frac{1}{K}\sum_{k=1}^{K}\mathbb{E}_{n,k}[m_{1}^{(2)}(Z,\tilde{\theta_{1}},\hat{\eta}_{1k})](\hat{\theta}_{1}-\theta_{1}^{*})
\end{align*}
and
\[
\sqrt{Nh^{3}}(\hat{\theta}_{1}-\theta_{1}^{*})=-\left[\frac{1}{K}\sum_{k=1}^{K}\mathbb{E}_{n,k}[m_{1}^{(2)}(Z,\tilde{\theta_{1}},\hat{\eta}_{1k})]\right]^{-1}\left(\frac{\sqrt{Nh^{3}}}{K}\sum_{k=1}^{K}\mathbb{E}_{n,k}[m_{1}^{(1)}(Z,\theta_{1}^{*},\hat{\eta}_{1k})]\right).
\]

In Step 1 and 2 below, we will show that

\[
\frac{1}{K}\sum_{k=1}^{K}\mathbb{E}_{n,k}[m_{1}^{(2)}(Z,\tilde{\theta_{1}},\hat{\eta}_{1k})]\overset{p}{\to}M_{1}
\]
and
\[
\frac{\sqrt{Nh^{3}}}{K}\sum_{k=1}^{K}\mathbb{E}_{n,k}[m_{1}^{(1)}(Z,\theta_{1}^{*},\hat{\eta}_{1k})]\overset{d}{\to}N(0,V_{1}),
\]
respectively. Hence, we can obtain the final result

\[
\sqrt{Nh^{3}}(\hat{\theta}_{1}-\theta_{1}^{*})\overset{d}{\to}N(0,M_{1}^{-1}V_{1}M_{1}^{-1}).
\]

\textit{Step 1.} Since $K$ is a fixed integer, which is independent of $N$, it suffices to show that for each $k\in [K]$,
\[
\mathbb{E}_{n,k}[m_{1}^{(2)}(Z,\tilde{\theta}_{1},\hat{\eta}_{1k})]\overset{p}{\to}M_{1}.
\]
Then we can show this convergence using the same argument in Step 1 in the proof of Theorem 2. 

\textit{Step 2.} Since $K$ is a fixed integer, which is independent of $N$, it is enough to consider the convergence of
$\mathbb{E}_{n,k}[m_{1}^{(1)}(Z,\theta_{1}^{*},\hat{\eta}_{1k})]$. Notice that

\begin{align*}
\mathbb{E}_{n,k}[m_{1}^{(1)}(Z,\theta_{1}^{*},\hat{\eta}_{1k})]&= \frac{1}{n}\sum_{i\in I_{k}}m_{1}^{(1)}(Z,\theta_{1}^{*},\eta_{10})+R_{2k}
\end{align*}
where
\[
R_{2,k}=\mathbb{E}_{n,k}[m_{1}^{(1)}(Z,\theta_{1}^{*},\hat{\eta}_{1k})]-\frac{1}{n}\sum_{i\in I_{k}}m_{1}^{(1)}(Z,\theta_{1}^{*},\eta_{10}).
\]
Then by triangular inequality, 
\[
\norm{R_{2,k}}\leq \frac{I_{1,k}+I_{2,k}}{\sqrt{n}},
\]
where

\[
I_{1,k}\equiv\norm{\mathbb{G}_{n,k}\left[m_{1}^{(1)}(Z,\theta_{1}^{*},\hat{\eta}_{1k})\right]-\mathbb{G}_{n,k}\left[m_{1}^{(1)}(Z,\theta_{1}^{*},\eta_{10})\right]},
\]
\[
I_{2,k}\equiv\sqrt{n}\norm{E_{P}\left[m_{1}^{(1)}(Z,\theta_{1}^{*},\hat{\eta}_{1k})\mid\left(W_{i}\right)_{i\in I_{k}^{c}}\right]-E_{P}\left[m_{1}^{(1)}(Z,\theta_{1}^{*},\eta_{10})\right]}.
\]
Two auxiliary results will be used to bound $I_{1,k}$ and $I_{2,k}$:

\[
\sup_{\eta_{1}\in\mathcal{T}_{N}}\left(E\left[\parallel m_{1}^{(1)}\left(Z,\theta_{1}^{*},\eta_{1}\right)-m_{1}^{(1)}\left(Z,\theta_{1}^{*},\eta_{10}\right)\parallel^{2}\right]\right)^{1/2}\leq\varepsilon_{N},\tag{A.1}
\]

\[
\sup_{r\in\left(0,1\right),\eta_{1}\in\mathcal{T}_{N}}\parallel\partial_{r}^{2}E\left[m_{1}^{(1)}\left(Z,\theta_{1}^{*},\eta_{10}+r\left(\eta_{1}-\eta_{10}\right)\right)\right]\parallel\leq\left(\varepsilon_{N}\right)^{2},\tag{A.2}
\]
where  $\mathcal{T}_{N}$
is the set of all $\eta_{1}=\left(\pi_{0},g_{10}\right)$ consisting of
square-integrable functions $\pi_{0}$ and $g_{10}$ such that 
\[
\parallel\eta_{1}-\eta_{10}\parallel_{P,2}\leq\varepsilon_{N},
\]
\[
\parallel \pi-1/2\parallel_{P,\infty}\leq1/2-\kappa,
\]
\[
\parallel \pi-\pi_{0}\parallel_{P,2}^{2}+\parallel \pi-\pi_{0}\parallel_{P,2}\times\parallel g_{1}-g_{10}\parallel_{P,2}\leq\left(\varepsilon_{N}\right)^{2}.
\]
Then by assumption, we have $\hat{\eta}_{1k}\in \mathcal{T}_{N}$ with probability $1-o\left(1\right)$.

To bound $I_{1,k}$, note that conditional on $\left(W_{i}\right)_{i\in I_{k}^{c}}$
the estimator $\hat{\eta}_{1k}$ is nonstochastic. Under the event
that $\hat{\eta}_{1k}\in\mathcal{T}_{N}$, we have 
\begin{align*}
E_{P}\left[I_{1,k}^{2}\mid\left(W_{i}\right)_{i\in I_{k}^{c}}\right]= & E_{P}\left[\parallel m_{1}^{(1)}(Z,\theta_{1}^{*},\hat{\eta}_{1k})-m_{1}^{(1)}(Z,\theta_{1}^{*},\eta_{10})\parallel^{2}\mid\left(W_{i}\right)_{i\in I_{k}^{c}}\right]\\
\leq & \sup_{\eta_{1}\in\mathcal{T}_{N}}E_{P}\left[\parallel m_{1}^{(1)}(Z,\theta_{1}^{*},\eta_{1})-m_{1}^{(1)}(Z,\theta_{1}^{*},\eta_{10})\parallel^{2}\mid\left(W_{i}\right)_{i\in I_{k}^{c}}\right]\\
= & \sup_{\eta_{1}\in\mathcal{T}_{N}}E_{P}\left[\parallel m_{1}^{(1)}(Z,\theta_{1}^{*},\eta_{1})-m_{1}^{(1)}(Z,\theta_{1}^{*},\eta_{10})\parallel^{2}\right]\\
= & \left(\varepsilon_{N}\right)^{2}
\end{align*}
by (A.1). Hence, $I_{1,k}=O_{P}\left(\varepsilon_{N}\right)$. To bound $I_{2,k}$, define the following function
\[
f_{k}\left(r\right)=E_{P}\left[m_{1}^{(1)}(Z,\theta_{1}^{*},\eta_{10}+r\left(\hat{\eta}_{1k}-\eta_{10}\right))\mid\left(W_{i}\right)_{i\in I_{k}^{c}}\right]-E\left[m_{1}^{(1)}(Z,\theta_{1}^{*},\eta_{10})\right]
\]
for $r\in[0,1)$. By Taylor series expansion, we have

\[
f_{k}\left(1\right)=f_{k}\left(0\right)+f'_{k}\left(0\right)+f''_{k}\left(\tilde{r}\right)/2,\text{for some }\tilde{r}\in\left(0,1\right).
\]
Note that $f_{k}\left(0\right)=E\left[m_{1}^{(1)}(Z,\theta_{1}^{*},\eta_{10})\mid\left(W_{i}\right)_{i\in I_{k}^{c}}\right]=E\left[m_{1}^{(1)}(Z,\theta_{1}^{*},\eta_{10})\right]=O(h^{2})$ by the calculation in Step 4 in the proof of Theorem 2.
Further, on the event $\hat{\eta}_{1k}\in\mathcal{T}_{N}$,
\[
\parallel f_{k}'\left(0\right)\parallel=\parallel\partial_{\eta_{1}}E[m_{1}^{(1)}(Z,\theta_{1}^{*},\eta_{10})]\left[\hat{\eta}_{1k}-\eta_{10}\right]\parallel=0
\]
by the orthogonality. Also, on the event $\hat{\eta}_{1k}\in\mathcal{T}_{N}$,
\[
\parallel f_{k}''\left(\tilde{r}\right)\parallel\leq\sup_{r\in\left(0,1\right)}\parallel f_{k}''\left(r\right)\parallel\leq\left(\varepsilon_{N}\right)^{2}
\]
by (A.2). Thus,
\[
I_{2,k}=\sqrt{n}\parallel f_{k}\left(1\right)\parallel=O_{P}\left(\sqrt{n}\left(\varepsilon_{N}\right)^{2}+\sqrt{n}h^{2}\right).
\]
Together with the result on $I_{1,k}$, we have 
\begin{align*}
\norm{R_{2,k}}\leq & \frac{I_{1,k}+I_{2,k}}{\sqrt{n}}\\
= & O_{P}\left(n^{-1/2}\varepsilon_{N}+\left(\varepsilon_{N}\right)^{2}+h^{2}\right)
\end{align*}
Hence, 
\[
\sqrt{Nh^{3}}\norm{R_{2,k}}=O_{P}(\sqrt{h^{3}}\epsilon_{N}+\sqrt{Nh^{3}}\epsilon_{N}^{2}+\sqrt{Nh^{3}}h^{2})=o_{P}(1)
\]
by the assumptions on the rate of convergence that $\epsilon_{N}=o((Nh^{3})^{-1/4})$ and $Nh^{7}\to 0$. Therefore,

\begin{align*}
\mathbb{E}_{n,k}[m_{1}^{(1)}(Z,\theta_{1}^{*},\hat{\eta}_{1k})]&= \frac{1}{n}\sum_{i\in I_{k}}m_{1}^{(1)}(Z,\theta_{1}^{*},\eta_{10})+o_{P}(1)\overset{d}{\to}N(0,V_{1}).
\end{align*}

\end{document}